\documentclass{aa}

\usepackage[]{aalongtable}
\usepackage{scalefnt}
\usepackage{graphicx}
\usepackage[english]{babel}

\begin{document}




\newbox\grsign \setbox\grsign=\hbox{$>$} \newdimen\grdimen \grdimen=\ht\grsign
\newbox\simlessbox \newbox\simgreatbox
\setbox\simgreatbox=\hbox{\raise.5ex\hbox{$>$}\llap
     {\lower.5ex\hbox{$\sim$}}}\ht1=\grdimen\dp1=0pt
\setbox\simlessbox=\hbox{\raise.5ex\hbox{$<$}\llap
     {\lower.5ex\hbox{$\sim$}}}\ht2=\grdimen\dp2=0pt
\def\simgreat{\mathrel{\copy\simgreatbox}}
\def\simless{\mathrel{\copy\simlessbox}}
\newbox\simppropto
\setbox\simppropto=\hbox{\raise.5ex\hbox{$\sim$}\llap
     {\lower.5ex\hbox{$\propto$}}}\ht2=\grdimen\dp2=0pt
\def\simpropto{\mathrel{\copy\simppropto}}

\title{A MUSE study of the inner bulge globular cluster Terzan~9: a fossil record in the Galaxy
 \thanks{Based on observations 
collected at the European Organisation for Astronomical Research in the Southern Hemisphere, 
Paranal, Chile, under ESO programme 097.D-0093 }}

%
\author{
H. Ernandes\inst{1,2,3}
\and
B.  Dias\inst{4,5}
\and
B. Barbuy\inst{1}
\and 
S. Kamann\inst{6}
\and
S. Ortolani\inst{7,8}
\and
E. Cantelli\inst{1}
\and
E. Bica\inst{9}
\and
L. Rossi\inst{10}
}
\offprints{H. Ernandes}
\institute{
Universidade de S\~ao Paulo, IAG, Rua do Mat\~ao 1226,
Cidade Universit\'aria, S\~ao Paulo 05508-900, Brazil
\and
UK Astronomy Technology Centre, Royal Observatory, Blackford Hill, Edinburgh, EH9 3HJ, UK
\and
IfA, University of Edinburgh, Royal Observatory, Blackford Hill, Edinburgh, EH9 3HJ, UK
\and
ESO, Alonso de Córdova 3107, Vitacura, Santiago de Chile, Chile
\and
Departamento de F\'{\i}sica, Facultad de Ciencias Exactas, Universidad Andr\'es Bello, Av. Fernandez Concha 700, Las Condes, Santiago, Chile
\and
Astrophysics Research Institute, Liverpool John Moores University, 146 Brownlow Hill, Liverpool L3 5RF, UK
\and
Universit\'a di Padova, Dipartimento di Fisica e Astronomia, Vicolo dell'Osservatorio 3, 35122 Padova, Italy 
\and
INAF-Osservatorio Astronomico di Padova, Vicolo dell'Osservatorio 5,
35122, Padova, Italy
\and
Universidade Federal do Rio Grande so Sul, Departamento de Astronomia, Av. Bento Gonçalves 9500, Rio Grande do Sul, Brasil
\and 
Centre for Astrophysics and Supercomputing, Swinburne University of Technology, Hawthorn, VIC 3122, Australia 
}

\date{Received ; accepted }

 
  \abstract
{Moderately metal-poor inner bulge globular clusters are relics of 
a generation of long-lived stars that formed in the early Galaxy.
Terzan~9, projected at 4\fdg12 from the Galactic center, is among the most central
 globular clusters in the Milky Way, showing an orbit which remains confined to the inner 1 kpc.
}
{Our aim is the derivation of the cluster's metallicity, together
with an accurate measurement of the mean radial velocity. In the literature, metallicities in the
range between $-$2.0$<$[Fe/H]$<$$-$1.0 have been estimated for Terzan 9 based on
 color-magnitude diagrams and CaII triplet (CaT) lines.
}
{Given its compactness, Terzan 9 was observed using the Multi Unit Spectroscopic Explorer (MUSE)
at the Very Large Telescope.
The extraction of spectra from several hundreds of individual stars allowed us to 
derive their radial velocities, metallicities, and [Mg/Fe]. 
The spectra obtained with MUSE were analysed through
full spectrum fitting using the ETOILE code.}
{We obtained a mean metallicity of [Fe/H]$\approx$-1.10 {$\pm$0.15}, a heliocentric radial velocity of v$^{\rm h}_{\rm r}$ = 58.1  $\pm$ 1.1 km s$^{-1}$  , and
a magnesium-to-iron [Mg/Fe] = 0.27 $\pm$ 0.03. 
The metallicity-derived character of  
Terzan 9 sets it among the family of the moderately metal-poor Blue Horizontal Branch clusters  HP~1, NGC 6558, and NGC 6522.}
{}
\maketitle


\section{Introduction} 

Globular clusters in the central parts of the Galaxy are among the oldest
extant stellar populations in the Milky Way (e.g. Barbuy et al. 2018a;
Kunder et al. 2018). 
Terzan~9 is a very compact cluster located at 4\fdg12 and 0.7 kpc
(Bica et al. 2006) from
the Galactic center, which is, thus, in the inner bulge volume, 
and it is among the globular clusters closest to the Galactic center. 
Terzan~9  appears to show a blue horizontal branch (BHB)
 in the ground-based color-magnitude diagrams (CMDs) by Ortolani et al. (1999).
The clusters identified with a moderate metallicity and a BHB
are very old as deduced from proper-motion cleaned 
color-magnitude diagrams (CMDs) for example for
NGC 6522 and HP~1 (Kerber et al. 2018, 2019). 
A proper-motion cleaned CMD for Terzan 9 is presented in
Rossi et al. (2015), with the cluster proper motions derived. 
Orbit calculations by Pérez-Villegas
et al. (2018) reveal that Terzan 9 remains confined within
1 kpc of the Galactic center with an orbit co-rotating with the bar,  
it has a bar shape in the (x  - y) projection, and a boxy shape in 
(x - z ), which indicates that these clusters are trapped by the bar. 
With absolute 
proper motions from Gaia DR2, a new orbital analysis was carried out (Pérez-Villegas et al. 2019) using a Monte Carlo method to take into account 
the effect of the uncertainties 
in the observational parameters.
These calculations  confirm that Terzan 9 belongs to the bulge globular
cluster group and that most of its probable orbits follow the bar. 
 Since the bulge clusters are typically old, they were
probably formed early in the Galaxy and were later
trapped by the bar (see also Renzini et al. 2018).
As a matter of fact, the bar should have formed at about 8$\pm$2 Gyr
 ago, according to Buck et al. (2018).

A metallicity of [Fe/H]$\sim$-2.0 is deduced by Ortolani et al.
(1999) and [Fe/H]$\sim$-1.2 by Valenti et al. (2007)
  from CMDs. Armandroff \& Zinn (1988) obtained [Fe/H]=-0.99
  from measurements of CaT lines.
  V\'asquez et al. (2018)
(ESO  proposal 089.D-0493) 
measured the CaT lines for six stars and obtained [Fe/H]$\sim$-1.08,
-1.21, and -1.16 following calibrations from
Dias et al. (2016), Saviane et al. (2012), and
V\'asquez et al. (2015), respectively. In the compilations by
Harris (1996, Edition of 2010)\footnote{www.physic.mcmaster.ca/$\sim$harris/mwgc.dat} and
Carretta et al. (2009), metallicities of
[Fe/H] = -1.05 and -2.07 are respectively reported.
 Given that spectroscopic results are more reliable for metallicity
derivations, it appears that a value of around [Fe/H]$\sim$-1.0
should be preferred. The aim of this work is to obtain the metallicity
derivation for Terzan~9, together with its radial velocity. 
The coordinates and typical photometric parameters for Terzan~9
 are reported in Table 1. 
 

\smallskip

\begin{table}[!h]
\caption{Terzan 9: data from literature. 
References: (1) Rossi et al. (2015), (2) Bica et al. (2006), (3) Ortolani et al. (1999), (4) Harris (1996, 2010 Edition).}
\centering 
\begin{tabular}{c|c} 
\hline
\hline
RA J2000  & 18 01 38.80   \\
\hline
DEC J2000  & -26 50 23.0   \\
\hline
l($^{\circ}$)  & 3.60   \\
\hline
b($^{\circ}$) & -1.99  \\
\hline
R$_{\rm Sun}$(kpc) & 7.7 (1,2) \\
\hline
R$_{\rm GC}$ (kpc) & 0.7 (2) \\
\hline
E(B-V) &   1.87 (2) \\
\hline
V$_{tip}$/V$_{HB}$ & 17.5/20.35 (3) \\
\hline
M$_{0\ V,t}$      & -3.71 (4) \\
\hline
\hline
\end{tabular}
\label{ter9} 
\end{table}

   \begin{figure}
   \centering
   \includegraphics[width=\hsize]{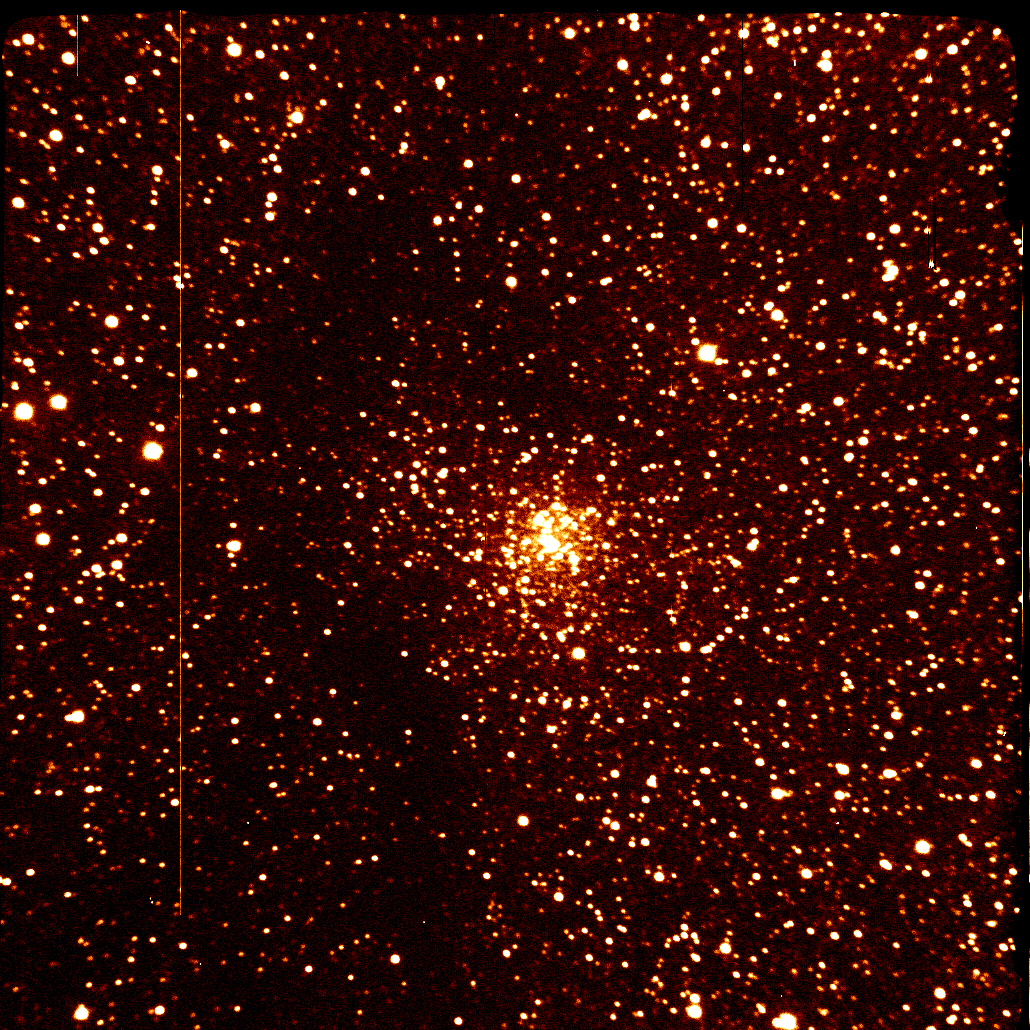}
      \caption{Terzan 9: I image of Terzan 9 obtained at NTT in 2012, 
with seeing of 0.5 arcsec. Size is 2.2x2.2 arcmin$^2$.
              }
         \label{image}
   \end{figure}

      \begin{figure}
   \centering
   \includegraphics[width=\hsize]{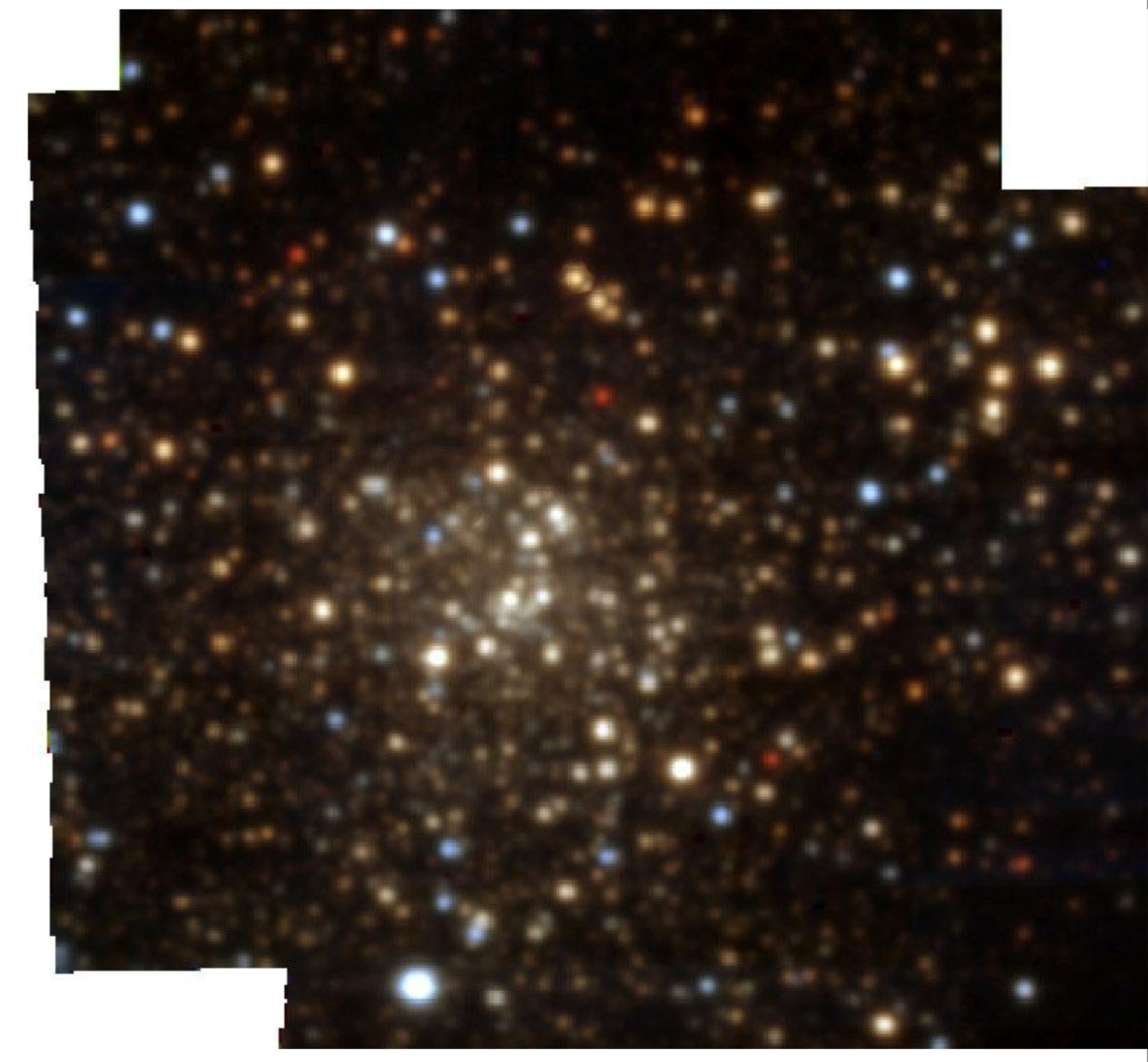}
      \caption{Terzan 9: composed image  in B, V and R, from 5 different pointings of Terzan 9 obtained with MUSE on Yepun in 2016. Size is 1.1x1.1 arcmin$^2$ , exposures with seeing from 0.51" to 1.01".
              }
         \label{imageMUSE}
   \end{figure}

The ETOILE code (Katz et al. 2011, Dias et al. 2015) is used
to derive the stellar parameters effective temperature, gravity,
metallicity, and [Mg/Fe] ratio for each sample star.
  This code corrects for radial velocity, compares the observed spectra of a sample star to all spectra from a grid of spectra,
and indicates which ones are the most similar. The procedure proved to work well, as demonstrated in Dias et al. (2015, 2016), where the method is applied
to 800 red giants in 51 globular clusters, observed with FORS2 at a similar resolution
as MUSE, which is of the order of R$\sim$2000 at 6000 {\rm \AA}.

In Sect. 2, we report our observations. In Sect. 3, the steps in data reduction are given.
Extraction of stellar spectra is described in Sect. 4. The analysis with the derivation of stellar
parameters and a discussion of results are presented in Sections 5 and 6. A summary  is
provided in Sect. 7.


\section{Observations}

Terzan~9 is faint and compact with a concentration factor of $c$ = 2.50 and
core radius $r_{\rm core}$ = 0.03$\arcmin$ 
(Harris 1996, updated in 2010), 
therefore the MUSE field of view (FoV) of 1'$\times$1'
appears suitable to locate and identify a large number of member stars.

The input list of stars was created from a combination of
the photometric observations of
Ortolani et al. (1999) with the Danish telescope in 1998  and
more recent observations with the NTT@ESO in 2012. 
The absolute calibration of the NTT 2012 data 
has been performed using our previous 1998 Danish data  (Ortolani et al. 1999).
About 800 stars in common between the Danish 1998 data and NTT 2012 have been
matched and checked in order to transform the instrumental NTT
magnitudes into the calibrated ones. Two almost linear relations  in magnitude and colors have
been found, with a residual slope, in a range within 0.01 mag,
possibly due to minor linearity deviations mostly at magnitudes
brighter than V$<$16. A simple offset has been applied then to the
instrumental magnitudes and colors.
The formal error of the transformation in  V and V-I is of about
0.025 magnitudes for both. The photometric error is dominated by linearity deviations at faint
magnitudes.
The V and I data were calibrated with the following
conversion coefficients:

\begin{equation*}
V_{calibrated} = V_{NTT2012} + 6.798\pm0.015
\end{equation*}
\begin{equation*}
(V-I)_{calibrated} = (V-I)_{NTT2012} + 1.77\pm0.02
\end{equation*}

These two sets of data were combined in Rossi et al. (2015) and used for
 proper-motion decontamination, making use of the 14 yr time difference
 between the 1998 and 2012 observations to have an optimized selection of member stars. 
We transformed the original data given in pixels in X,Y into 
right ascension and declination (RA,DEC) based on the NTT 2012 image.
The final coordinates are established by matching stars in common
with the Gaia Data Release 2 (Gaia Collaboration 2018).
The list of stars with their coordinates, along with their
V and V-I, are reported in Table \ref{Tabbasic}.
In Fig. \ref{image}, we show an I
image of Terzan 9 obtained at the NTT in 2012, with an excellent seeing
 of 0.5''.

The observations of the Terzan 9 field were conducted with the MUSE instrument installed
on the UT4 Yepun unit of the Very Large Telescope (VLT), with the Wide Field Mode, no-AO, standard
coverage (nominal mode WFM-NOAO-N). The FOV of MUSE in the Wide Field mode is 1'x1' per exposure. The total observing
time was 5 hours including overheads, that were distributed along 5 observation blocks with 3
exposures (one in the central field and 2 offsets) of 948 seconds each. Besides a rotation of
90$^{\circ}$, as recommended, and
offsets of $<$ 2 s in RA and of up to 18 s in DEC
were applied. Detailed information
about each exposure is given in Table \ref{logtable}. 
The MUSE datacubes were convolved with the transmission curves of the filters
Red, Green and Blue, resulting in three images. The color composite image in    B ( 4800 {\rm \AA}), V ( 5477 {\rm \AA}), and R ( 6349 {\rm \AA}). We note the Johnson-Cousins B filter overlaps only 22.77\% of the
MUSE wavelength coverage of 4800-9300 {\rm \AA}. That is the reason why
the B in the color composing image, Fig. \ref{imageMUSE} is centered in 4800 {\rm \AA} instead of 4353 {\rm \AA}.

\begin{table*}
\scriptsize
\caption{Observation log.}             
\label{logtable}      
\centering           
\begin{tabular}{c c c c c c c c }
\hline\hline  
cube name & date & exp. time & airm. start & airm. end & seeing start & seeing end & relative humidity \\
\hline
\noalign{\smallskip}
WFM\_Ter9\_OB1 exp1 & 2016-05-28T06:37:58 & 948 & 1.003 & 1.01  & 0.51 & 0.51 & 17.5 \\
WFM\_Ter9\_OB1 exp2 & 2016-05-28T06:55:37 & 948 & 1.011 & 1.023 & 0.56 & 0.62 & 17.5 \\
WFM\_Ter9\_OB1 exp3 & 2016-05-28T07:13:16 & 948 & 1.024 & 1.041 & 0.57 & 0.80 & 15.0 \\
\noalign{\smallskip}
WFM\_Ter9\_OB2 exp1 & 2016-05-28T07:39:44 & 948 & 1.053 & 1.078 & 0.72 & 0.53 & 17.0 \\
WFM\_Ter9\_OB2 exp2 & 2016-05-28T07:57:35 & 948 & 1.08  & 1.112 & 0.53 & 0.59 & 17.0 \\
WFM\_Ter9\_OB2 exp3 & 2016-05-28T08:15:43 & 948 & 1.115 & 1.155 & 0.59 & 0.87 & 15.5 \\
\noalign{\smallskip}
WFM\_Ter9\_OB3 exp1 & 2016-06-05T06:58:55 & 948 & 1.041 & 1.063 & 1.01 & 0.89 & 35.0 \\
WFM\_Ter9\_OB3 exp2 & 2016-06-05T07:16:45 & 948 & 1.065 & 1.093 & 0.97 & 0.91 & 35.0 \\
WFM\_Ter9\_OB3 exp3 & 2016-06-05T07:34:40 & 948 & 1.095 & 1.131 & 0.95 & 0.86 & 35.0 \\
\noalign{\smallskip}
WFM\_Ter9\_OB4 exp1 & 2016-06-09T03:35:40 & 948 & 1.116 & 1.083 & 0.77 & 0.63 & 4.5  \\
WFM\_Ter9\_OB4 exp2 & 2016-06-09T03:53:47 & 948 & 1.081 & 1.055 & 0.63 & 0.72 & 4.0  \\
WFM\_Ter9\_OB4 exp3 & 2016-06-09T04:12:05 & 948 & 1.053 & 1.033 & 0.73 & 0.67 & 4.0  \\
\noalign{\smallskip}
WFM\_Ter9\_OB5 exp1 & 2016-06-11T02:20:22 & 948 & 1.332 & 1.263 & 0.74 & 0.70 & 16.0 \\
WFM\_Ter9\_OB5 exp2 & 2016-06-11T02:38:13 & 948 & 1.259 & 1.201 & 0.70 & 0.71 & 16.0 \\
WFM\_Ter9\_OB5 exp3 & 2016-06-11T02:56:06 & 948 & 1.198 & 1.151 & 0.71 & 0.76 & 16.0 \\
\noalign{\smallskip}
\hline\hline                  
\end{tabular}
\end{table*}

\section{Data reduction}

\subsection{Individual exposure reduction and sky subtraction}
The individual exposures were reduced using the MUSE instrument pipeline v2.0.1 under the
Reflex environment.

Since the field is highly crowded and reddened due to its location at low latitudes in the Galactic bulge, the 
sky subtraction must be carefully conducted. The MUSE pipeline gives the
choice of the fraction of the FOV to be considered as sky. After some
tests, we chose 6\% based on the generated mask area, zones affected by sky contamination on the stellar spectra, and sky spectra comparisons assuming different fractions. For fractions much higher than 6\%, the sky spectrum shows
absorption lines of unresolved faint stars at redder wavelengths. 
For fractions much below 6\% the final sky spectrum is not representative
of the whole FOV, which implies that some sky-subtracted stellar spectra
still show some sky emission lines.

\begin{figure}
   \centering
   \includegraphics[width=\hsize ]{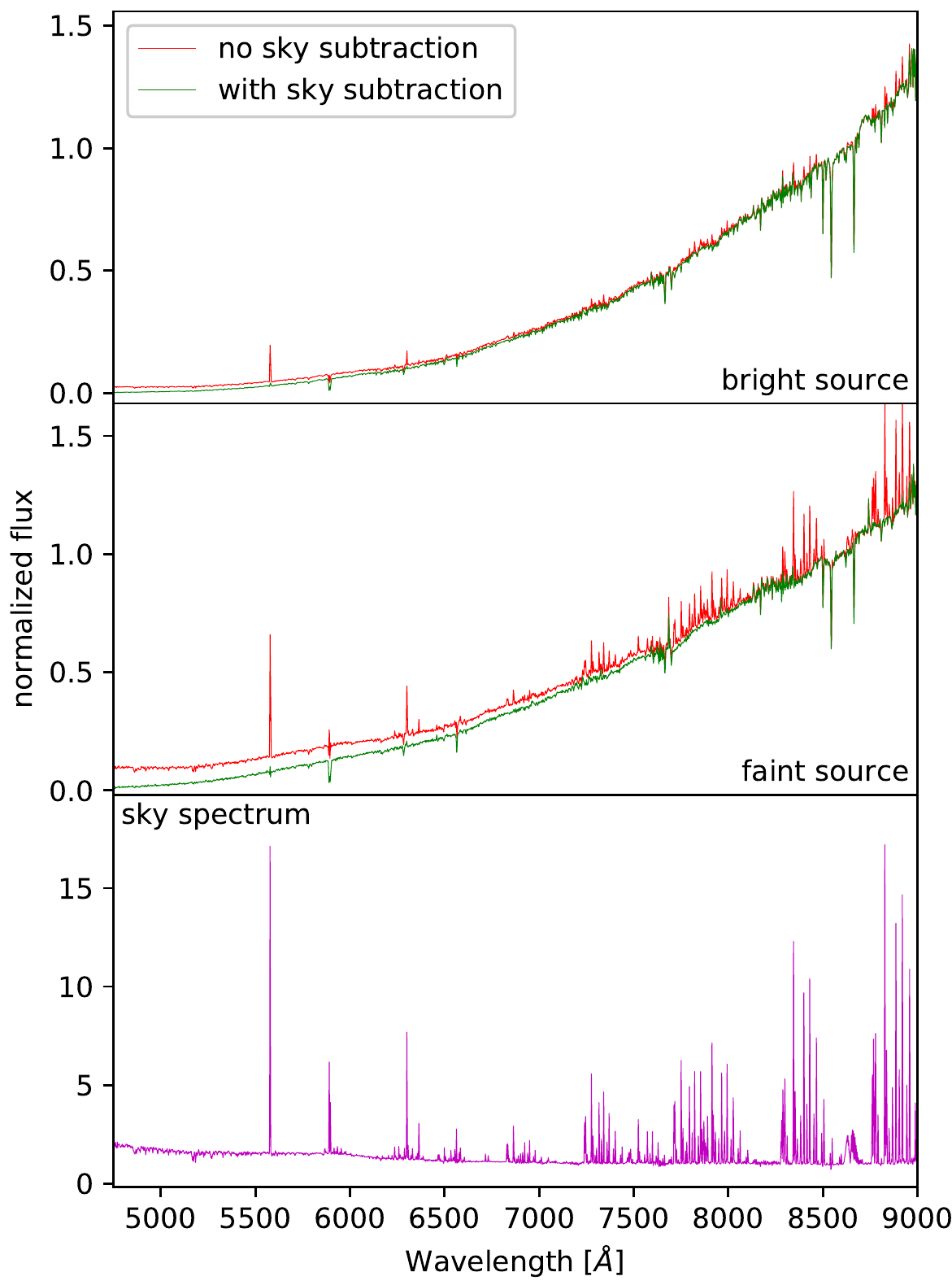}
      \caption{Comparison between sky contribution to bright and faint stellar sources.
       Spectra are normalized to their flux at 8604 {\rm \AA}. 
              }
         \label{scattersky}
   \end{figure}

\begin{figure}
   \centering
   \includegraphics[width=\hsize]{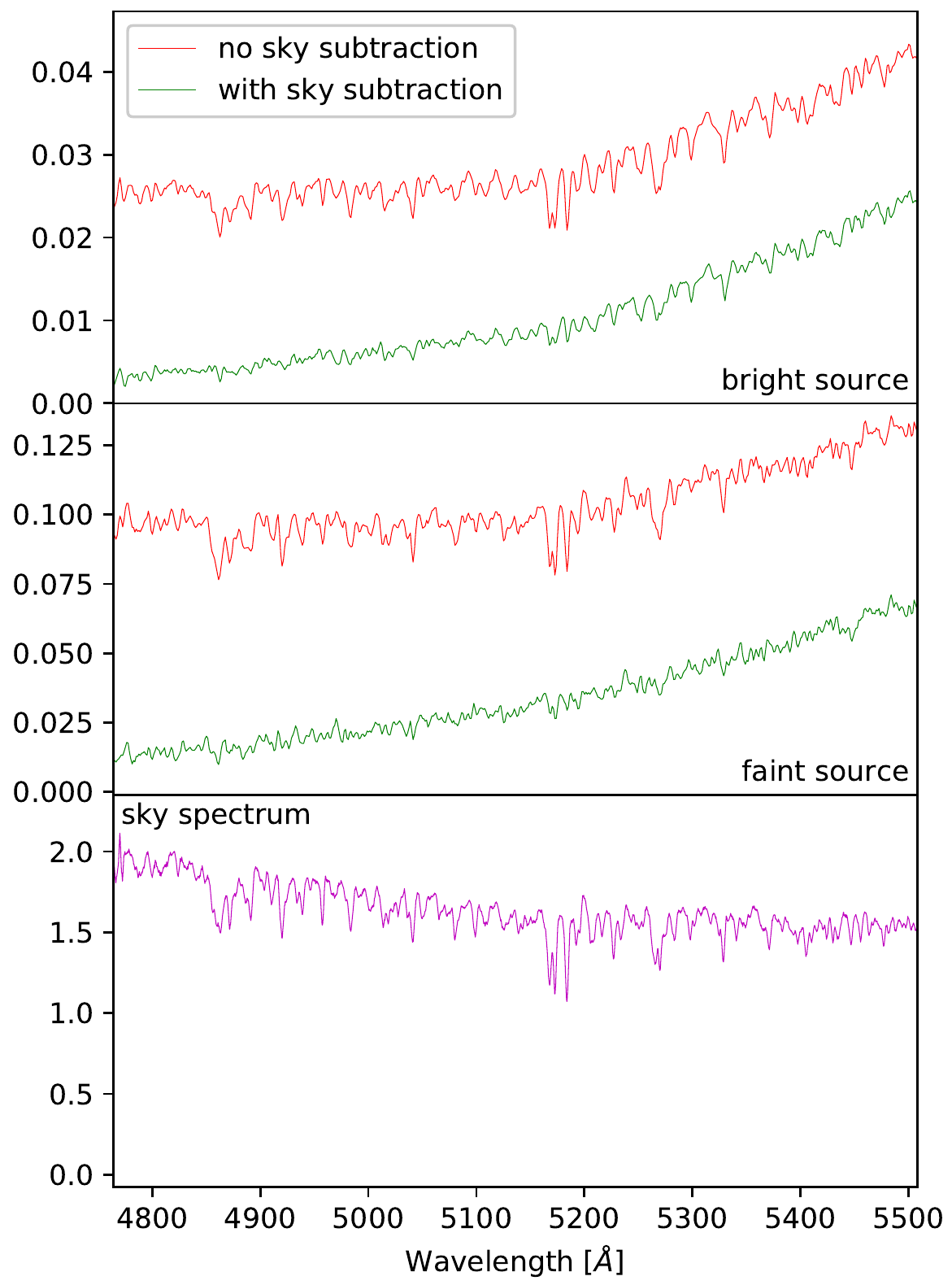}
      \caption{Comparison between sky contribution to bright and faint stellar sources, normalized to their flux at 8604 {\rm \AA}. Scattered stellar light absorption lines can be seen in the sky spectrum, and its subtraction from the actual star spectrum preserves the true line profile.
              }
         \label{scattersky-blue}
   \end{figure}

The subtraction method used is simple,
demonstrating a slight improvement in the signal-to-noise (S/N).
Figures \ref{scattersky}  (full wavelength range)
and \ref{scattersky-blue}  (zoom on the bluest range) show the effect of sky subtraction.
The faint sources are more affected by the sky than the bright ones because of their flux level being lower and closer to the sky level.

An  example of spectra from different datacubes are shown in Fig. \ref{spec274} for a sample star. 

\subsection{Datacube combination}

The combination of the exposures was done by observation block (OB) using the most recent version (v2.1.1) of the MUSE instrument pipeline  available only in the Gasgano environment.
Gasgano's interface allows for a quick assignment of frames to specific recipes
and easy parameter manipulation, together with a processing request pool, so
it is convenient for doing tests and requesting different datasets.
We combined the three exposures of each OB to end up with five final cubes. The combination of
all OBs was not done in the same way 
because they were observed in different conditions. The final stellar spectra
correspond to the combination of the extracted 1D spectra of each star from the five cubes.


During the combination, several tests were carried out. The most influential parameter was the
resampling method in building the combined cube.
The MUSE pipeline default method is "drizzle" and comparisons between this method, along with other
complex methods "renka" and "lanczos," were performed. The renka method showed the best spatial resolution and image coverage among the three.
We performed some tests with the renka resample method  to find the critical radius $cr$ value that optimizes the S/N of the extracted spectra, starting with the default value $cr = 1.25$. We noticed that the S/N increases for $cr < 0.1$ and that the line spread function (LSF) starts to degrade if we adopt $cr < 0.03$, therefore we chose $cr = 0.03$ to optimize the S/N of the extracted spectra without degrading the LSF. We also note that the reconstructed images using $cr = 0.03$ reveal fainter stars with a stable PSF and higher S/N, delivering a better result than with the default parameters.

In addition, there were  three other, simpler resampling methods: nearest, quadratic, 
and linear. A comparison between these three and the more complex methods discussed above 
showed that the linear method achieved even better results than `renka' both in terms of the S/N ratio, and the spectral and spatial resolution. Our final resampling was done using the linear method.
All of the comparisons were made visually with different source brightness in the regions of the \ion{Mg}{I} triplet, H$\alpha$ and \ion{Ca}{II} triplet, as well as the spatial resolution and PSF quality, using DS9.

For each of the final cubes, 2D images were created by multiplying the cube by  filters transmission curves  available in the pipeline: Johnson B, V, Cousins R, I,  and a few HST-ACS filters. 
These images were used to generate
color-magnitude diagrams (CMDs) and select Red Giant Branch (RGB) stars to
be cross-matched with our
previous catalogue.


\subsection{Extraction of Stellar Spectra}

 To extract the data from the MUSE datacubes, we employed the 
PampelMUSE\footnote{\url{https://gitlab.gwdg.de/skamann/pampelmuse}} code (Kamann et al. 2013) which is specific to stellar spectra extraction in crowded fields of data cubes such as MUSE. This software aptly deals with the observation of a densely populated stellar field such as
 a globular cluster.   One challenge is the seeing-limited angular resolution of the instrument.  A single object is represented by a point spread function (PSF),
and the stellar field is a sum of many overlapping PSFs. Even in cases of heavily blended regions, the objects can be recovered using a PSF model if the distance between two neighbor stars is larger than $0.3\times$FWHM.
 
  This code written in python executes many tasks.  In a simplified picture, a datacube is a sum of layers in wavelength of the image.
 A spaxel contains the entire spectrum, hence contributes to all layers.
This method consists in analysing the datacube,
 layer by layer, performing PSF photometry individually on each layer. In the end all photometric solutions for each layer are combined, 
building  spectra for each of the objects.
  
 In order to get the spectra of sources of interest from the datacube, it is needed
to provide an input catalogue with the position and magnitude of these objects, or else a selection by hand on the image.
The coordinates are identified in the list of stars from the NTT 2012 observations (Sect. 2), 
and the proper motion cleaned CMDs by  Rossi et al. (2015). The code locates the
stars through a PSF fitting;
 a degree of confidence  is assigned to each object, that can then be resolved in the crowded stellar field, and the spectra
 to be extracted.
 
  To find a PSF in  a crowded field,  
the program selects a number of relatively isolated objects and fits to them an
 analytical function. Then an Hermitian of order two is used to smooth the PSF
 parameters as a function of wavelength.
  

 
\begin{figure}[!htb]
\begin{center}
\includegraphics[width=3.1in]{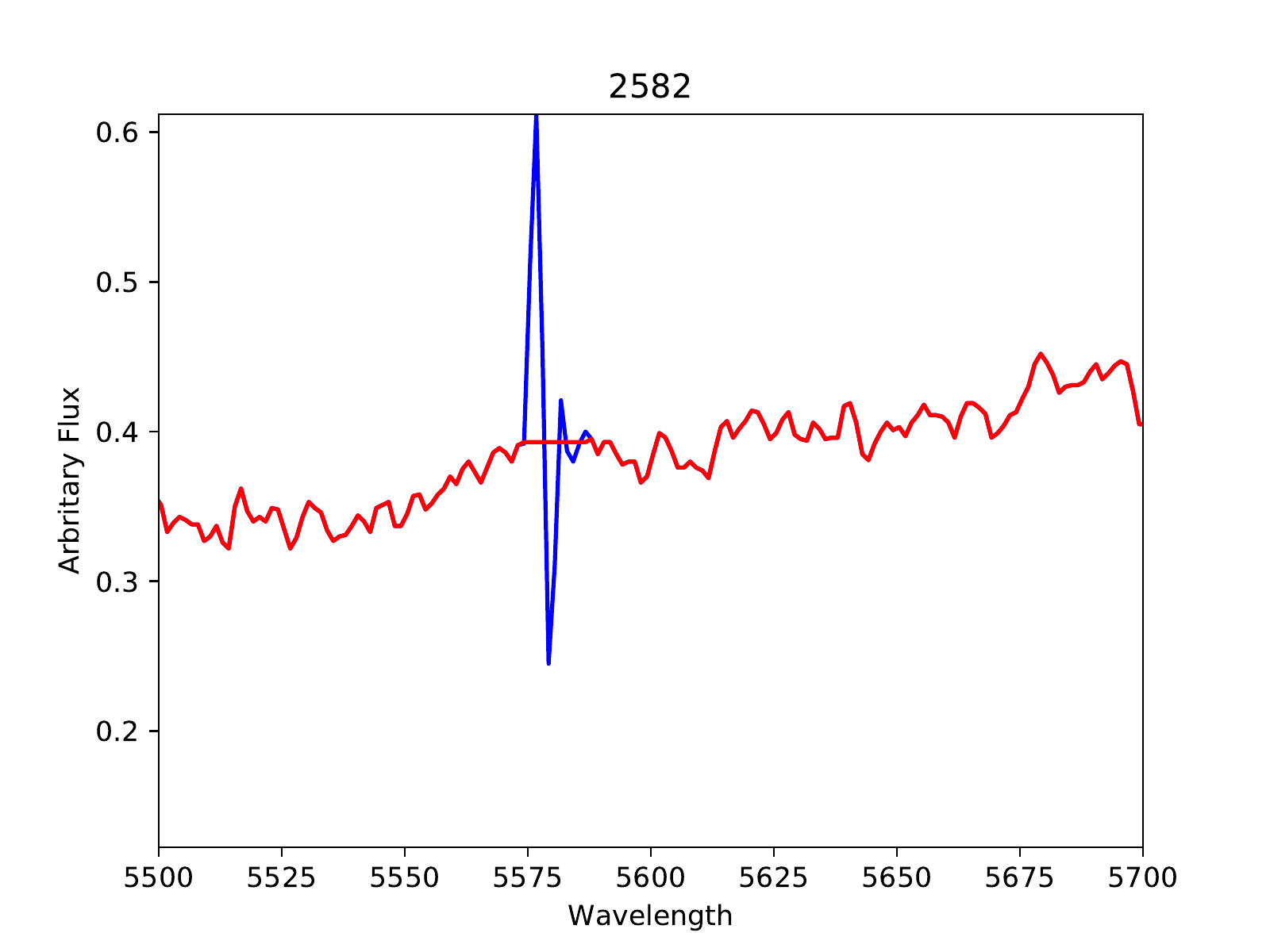}
\caption{Emission line subtraction from spectrum of star 2582.}
\label{cosmosub}
\end{center}
\end{figure}

The last step in the data handling before the analysis is the removal of emission lines and non-stellar features left behind in the previous steps. These lines could introduce noise to the results in the minimum distance method which is the basis of the code ETOILE.
The elimination of emission lines was made using a python code, which identifies the lines
and cuts them in a region between their two edges, as illustrated in Fig. \ref{cosmosub}. 
We proceeded with the elimination of
the emission line [O I] 5577.338 {\rm \AA} (Osterbrock et al. 1996) from all sample spectra. A future version of ETOILE may have the option of masking out undesired regions, such as those with their emission lines remaining after the cosmic ray cleaning and sky subtraction.

\begin{figure}[!htb]
\begin{center}
\includegraphics[width=3.5in]{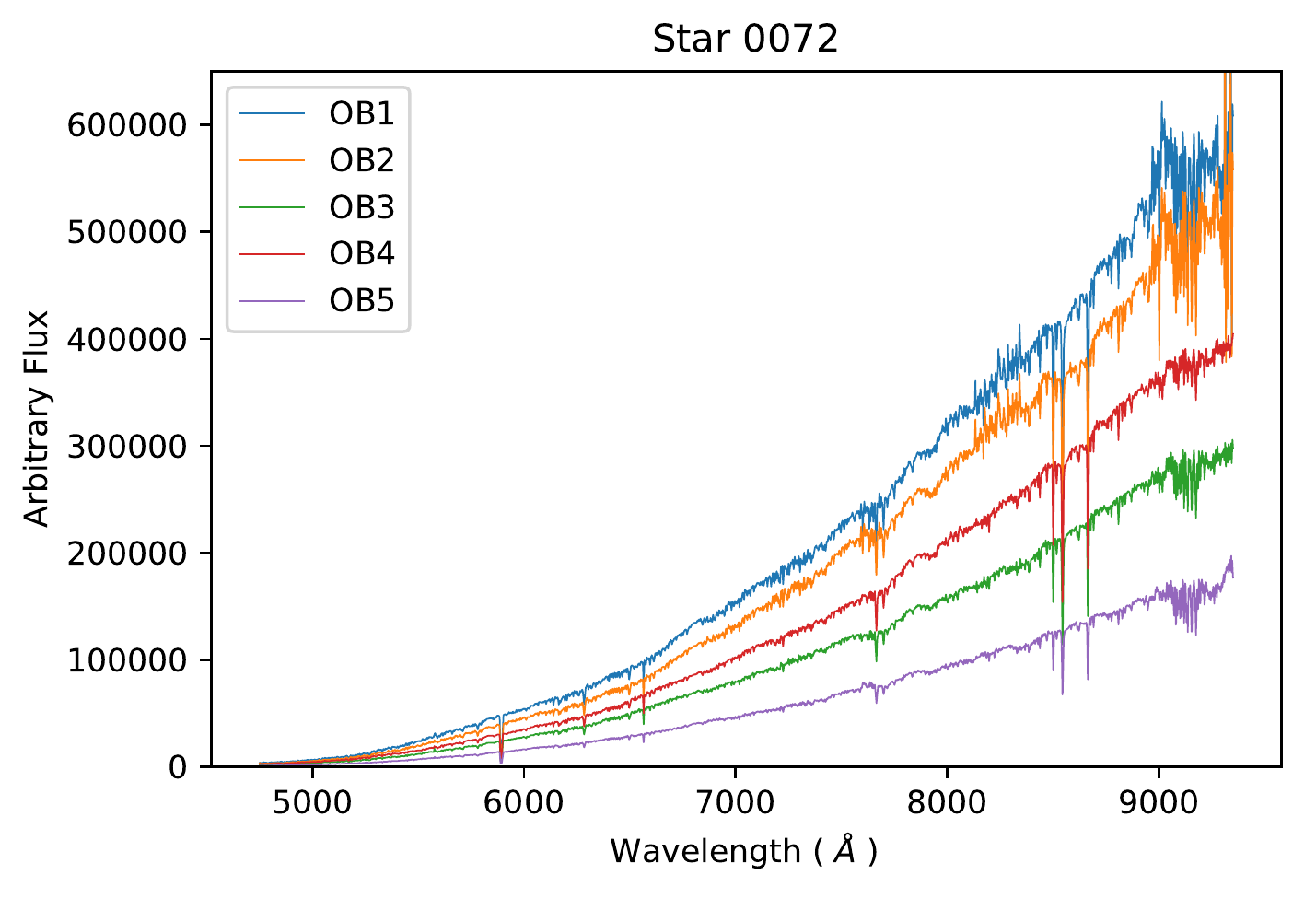}
\caption{Spectra of Terzan 9 star id 0072 from different data cubes obtained with PampelMUSE.}
\label{spec274}
\end{center}
\end{figure}

 Finally, the extracted spectra for each star observed in different nights were combined to get a final 1D spectrum with higher S/N for each star. The combination is done following these steps: Fig. \ref{spec274} shows the difference in flux levels in the spectra of a same star observed on different nights with the same exposure time but different weather conditions. We accounted for the difference in flux by 
 normalizing them at 5000 {\rm \AA} and adding up all with no airmass-based or
 S/N-based weight, given that for the same star, there is little variation in S/N. S/N$\sim$110 for V$\sim$17, and S/N$\sim$90 for V$\sim$20. 
 All S/N values are given in Table \ref{Tabbasic}.

\section{Analysis}

We derived atmospheric parameters via full spectrum fitting
with the ETOILE code (Katz et al. 2011). This method is very robust in finding the absolute minimum
in a $\chi^2$ map (e.g. Recio-Blanco et al. 2014; Jofre et al. 2018).
The code written in C is a modified version of the HALO (Cayrel et al. 1991) and TGMET (Katz et al. 1998) codes, which is obtained by
changing the main four procedures:  a) the sample star spectrum is compared with the full list of reference spectra, b) the input data are in ascii format, 
c) the target spectrum does not need to be normalized or calibrated in absolute flux, and d) no input parameters are given.
More details on the method for extracting the
fundamental stellar parameters (T$_{\rm eff}$, log~$g$, [Fe/H]) from the spectra are given
in Katz et al. (1998, 2001). In the original code, high resolution spectra of 2000 stars
obtained with the ELODIE spectrograph, as presented in Katz et al. (2011), were adopted
as reference.

Dias et al. (2015, 2016) implemented two other grids of spectra suitable for the analysis of medium-resolution spectra in the wavelength range
4600-5600 {\rm \AA}: the synthetic spectra
by Coelho et al. (2005, hereafter Coelho05) and the MILES grid of observed spectra
(S\'anchez-Bl\'azquez et al. 2006). 
We implemented a wavelength-extended version to be run with the Coelho05 library, 
encompassing the range 3000-18000 {\rm \AA} 
that covers the region of the MUSE spectra 4800-9300 {\rm \AA}, and it was used in different ways, as explained below.

\begin{figure}[!htb]
\begin{center}
\includegraphics[width=3.5in]{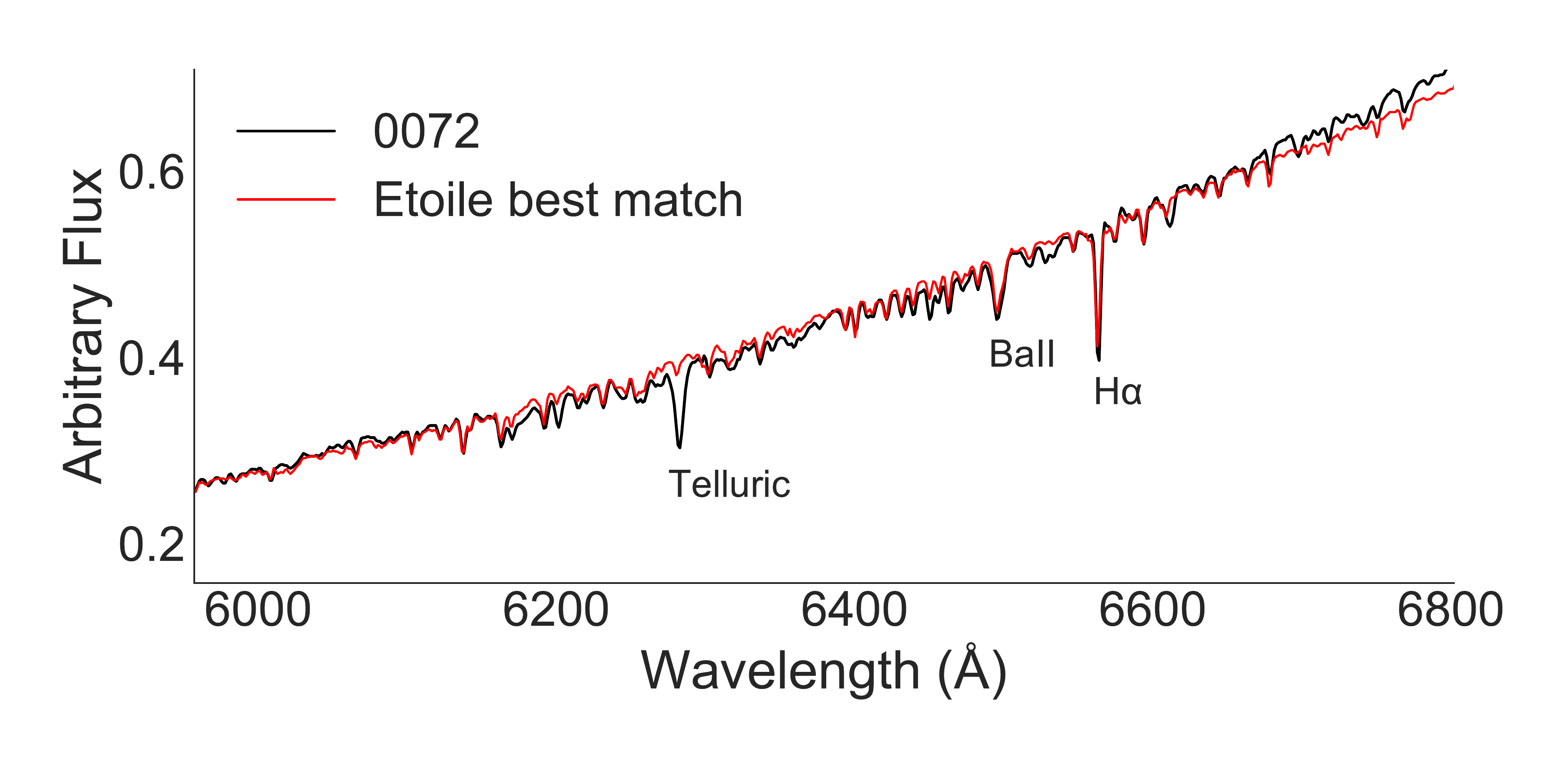}
\includegraphics[width=3.5in]{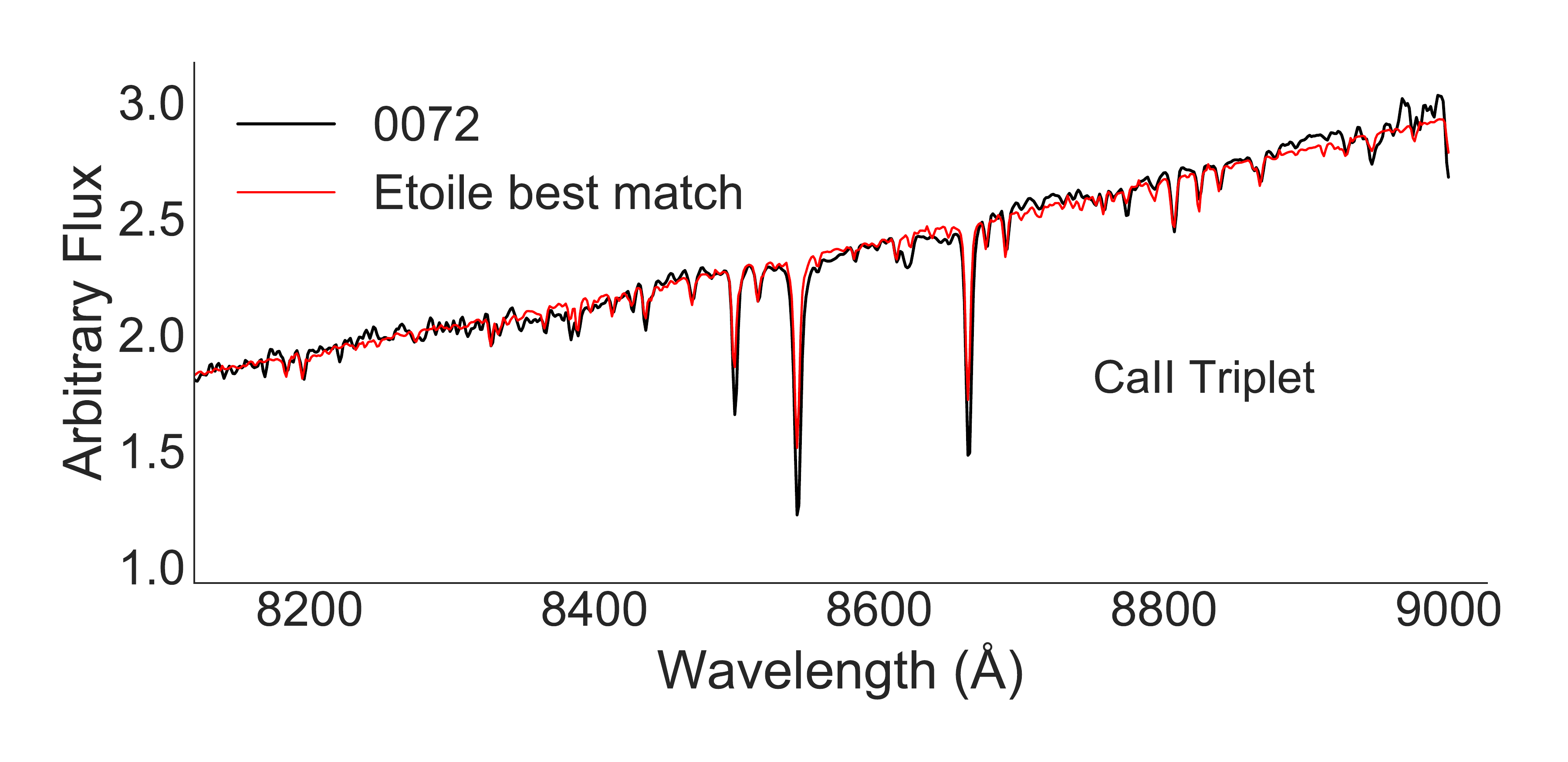}
\caption{Fit obtained with ETOILE for star 0072  with a good S/N =129.80. Upper panel: region 6000- 6800 {\rm \AA} 
where the strongest lines (telluric feature at 6282 {\rm \AA}, BaII 6496.9 {\rm \AA}
and H$\alpha$) are indicated; lower panel: calcium triplet region.}
\label{compared}
\end{center}
\end{figure}

In summary, the ETOILE\ code compares the observed
spectrum to a list of reference spectra, either observed or synthetic, and finds the most similar ones through a least square of Euclidean distance measure.
An example of a fit to a sample spectrum is given in Fig. \ref{compared}.

\subsection{Sample extraction and radial velocities}

\begin{figure}[!htb]
\begin{center}
\includegraphics[width=3.7in]{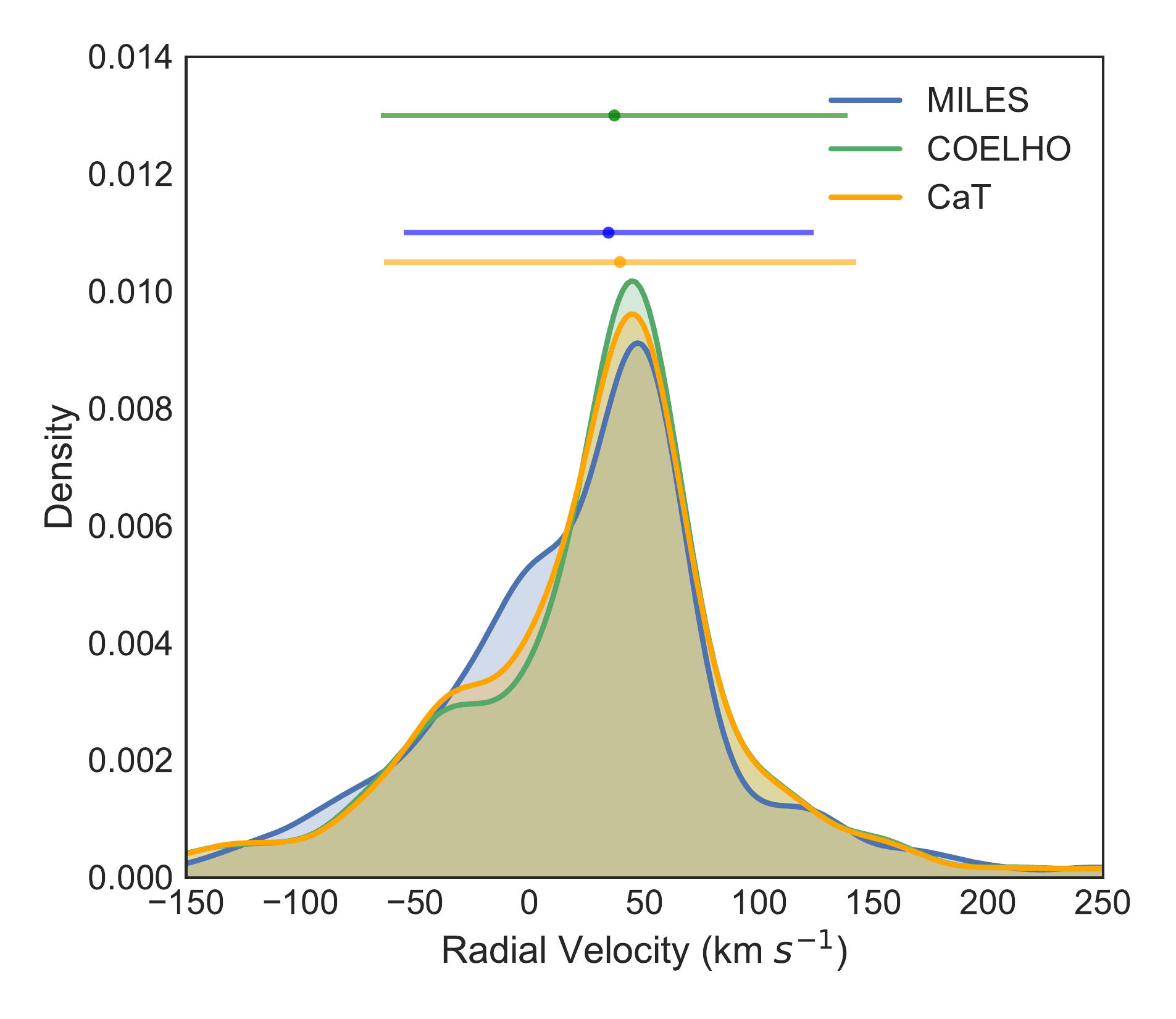}
\caption{Histograms of radial velocities obtained in the cases: 
green distribution: MILES library in the
range 4860 - 5600 {\rm \AA};  blue distribution: Coelho05 library in the range 4860 - 9000 {\rm \AA};
and yellow distribution: Coelho05 library in the CaT region at 8400 - 8750 {\rm \AA}.}
\label{vrall}
\end{center}
\end{figure}

We were able to extract and combine spectra from the five data cubes for 614 stars. After a selection based on S/N (S/N $\geq $85) of all final spectra, 90 of them were retained for analysis. The choice of this high S/N was due to
better reliability in the parameter derivation.
The ETOILE code was run for these spectra in order to derive their stellar parameters. 
The code first corrects for radial velocity (v$_{r}$) through cross-correlation with a template spectrum 
from the library in use. 
In the present case, we used the MILES library in the wavelength range 4600 - 5600 {\rm \AA}, the 
synthetic Coelho05 library in the full MUSE range  4860  - 9300 {\rm \AA}, and
in the region of the CaII triplet (CaT) 8400 - 8750 {\rm \AA}.
The use of these different libraries and wavelength regions has
shown that the most reliable method  to derive radial velocities 
is the comparison of the sample spectra with the synthetic spectra in the CaT region.
We concluded this from inspecting a series
of spectra from the full initial sample and comparing them individually to reference spectra,  verifying
the wavelength region with that particular radial velocity value.
The results are shown in Fig. \ref{vrall} as smoothed histograms 
of radial velocities obtained in the three cases described above.

\begin{figure}[!htb]
\begin{center}
\includegraphics[width=3.7in]{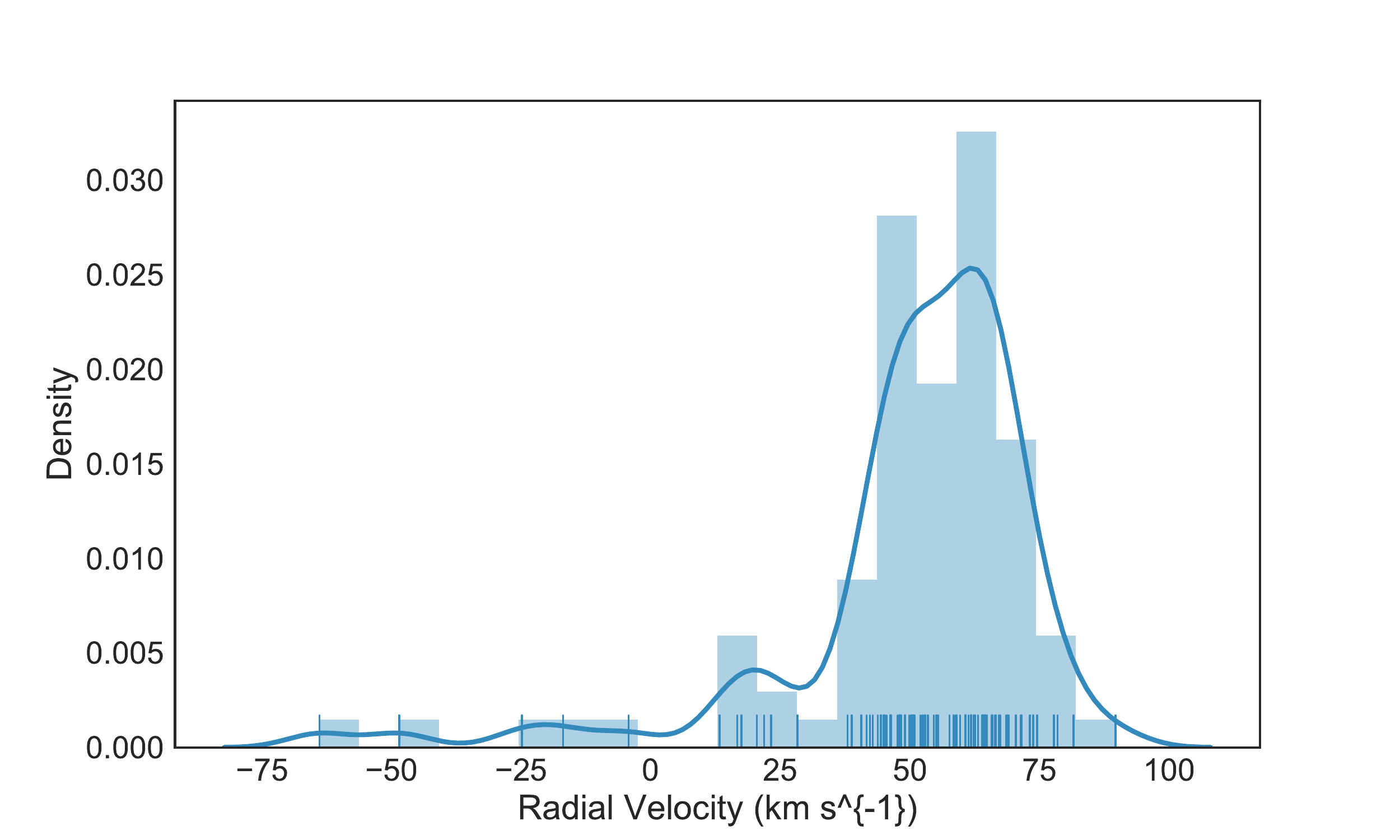}
\caption{Smoothed histogram of radial velocities obtained with the
 Coelho05 library in the CaT region at 8400 - 8750 {\rm \AA}. A kernel density estimation (KDE) gaussian fitting the main peak of radial velocity distribution
 is overplotted.}
\label{vrcat}
\end{center}
\end{figure}

Fig. \ref{vrcat} shows the radial velocity  distribution using the CaT region
analysed through the Coelho05 library, for the 90 selected stars. A
gaussian fit  results in a mean radial velocity value  of v$_{\rm r}$ = 
49.7 km s$^{-1}$ and a sigma of 22 km s$^{-1}$.
The 
mean heliocentric radial velocity is
v$^{\rm h}_{\rm r}$ = 58.1 km s$^{-1}$. 
The radial velocity of 
  v$^{\rm h}_{\rm r}$ = 71.4 $\pm$0.4km s$^{-1}$  from six stars by
V\'asquez et al. (2018) is compatible with the present value within uncertainties.
A comparison with two stars in common with V\'asquez et al. (2018)
is reported in Table \ref{vasquez}, showing excellent agreement
in terms of radial velocities.
In conclusion, we suggest that the present value is more accurate given the larger sample of stars taken into account.

For these two stars in common, 
the metallicities from the present work,
derived with ETOILE and from CaT with the same method as V\'asquez et al. (2018),
 that is, by applying their Equation 5 for the metallicity scale by Dias et al. (2016) and their reported values, given in Table \ref{vasquez}, show
good agreement within uncertainties. The full explanation on how the metallicities are calibrated is given on Sect. 4.4

Finally, the spectra are corrected for the adopted results of radial velocity, which are reported in Table \ref{Tabfinal}.

\begin{table}
\scriptsize
\caption{Comparison of radial velocity and metallicity for two stars in common with V\'asquez et al. (2018, V+18). 
The metallicity from V+18 adopts the metallicity scale by Dias et al. (2016).}             
\label{vasquez}      
\centering    
\scalefont{0.8}
\begin{tabular}{c c c c c@{} c@{} c@{} }
\hline\hline  
\noalign{\smallskip}
ID & ID$_{V+18}$ & v$_{r}$ &  v$_{r}$(V+18) &  [Fe/H] & \phantom{-}\phantom{-}[Fe/H] &  \phantom{-}\phantom{-}[Fe/H]$_{V+18}$ \\
 &  & km.s$^{-1}$ &  km.s$^{-1}$ & ETOILE  & CaT &   \\
\hline
\noalign{\smallskip}
1322   &  1\_399   & 75.9$\pm$1.1  &  74.8$\pm$0.7 & -1.52 & -1.14 & -1.25 \\
1378   &  1\_745   & 60.8$\pm$1.1  & 61.9$\pm$0.6  & -1.23 & -1.34 & -1.26 \\
\noalign{\smallskip}
\hline\hline                  
\end{tabular}
\end{table}


\subsection{Coordinates and proper motions}

The X,Y position of stars in the NTT image used to identify the stars
in the MUSE data, were transformed to right ascension (RA) and declination (DEC)
and matched with the Gaia Data Release 2 (DR2, Gaia collaboration 2018) coordinates, therefore the
coordinate values reported in Table \ref{Tabbasic} have  a high astrometric precision.   For the list of 90 selected stars,  Gaia data are available.

\begin{figure}[!htb]
\begin{center}
\includegraphics[width=3.5in]{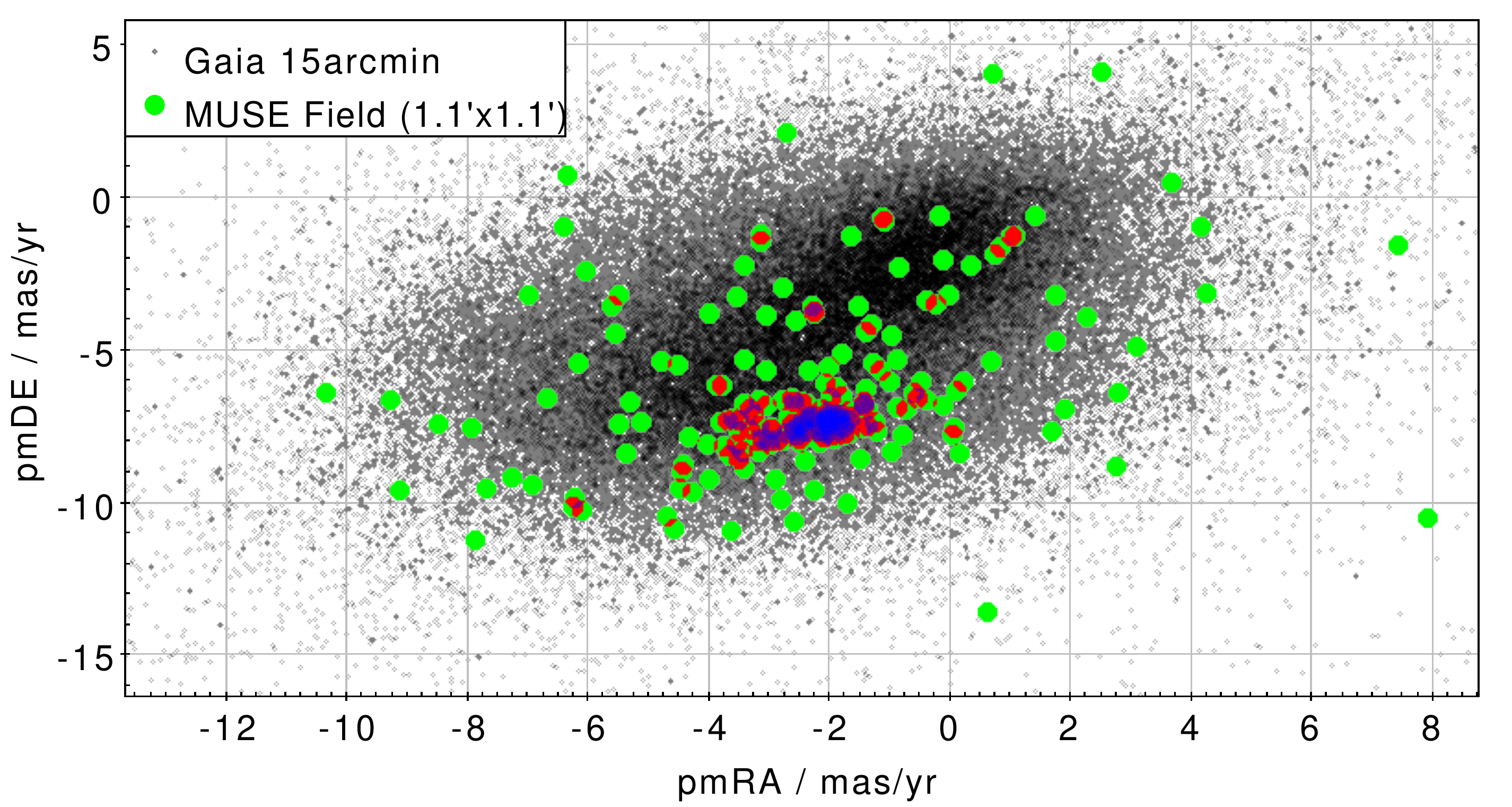}
\caption{Proper motions from Gaia. Symbols: gray dots: Gaia stars contained within
a radius of 15 arcmin from the cluster center; Green dots: Gaia stars in a density representation enclosed in the MUSE field (1.1'$\times$1.1'). The clustering of stars from Terzan 9 can be seen in  red and blue, where blue is the densest part.}
\label{propermotion1}
\end{center}
\end{figure}

In the MUSE field, there are 371 stars in the Gaia data that
are shown in Fig. \ref{propermotion1}, where
 we see  a clear cluster, seen as the feature highlighted in blue.
 Among the 371 Gaia stars, we identified  236 stars with proper motion (PM) information.
 For this sample, the mean proper motion values derived are:
  pmRA = -2.212$\pm$ 0.0851 mas/yr, and pmDE = -7.425$\pm$ 0.0851 mas/yr, in good
 agreement with derivations by P\'erez-Villegas et al. (2019) of
   (-2.314$\pm$0.108, -7.434$\pm$0.068) mas/yr
and  (-2.225$\pm$0.038,-7.492$\pm$0.029) mas/yr from Vasiliev (2018).
 Note that the PM value derived uses 236 stars from Gaia which are present 
in the MUSE field. The values are the same as for the 90 selected member stars,
as made evident in the corner plot given in Fig. \ref{cornerplot}.
We note that the previous values
 by Rossi et al. (2015) of (0.0$\pm$0.38,-3.07$\pm$0.49) were different
from these data, which are more accurate.

In order to identify a final list of member stars, we selected stars from
their radial velocity of v$_{\rm r}$ = 58.1$\pm$1.1 km s$^{-1}$, combined
with proper motions of pmRA = -2.21$\pm$0.10 and pmDEC = -7.42$\pm$0.07.
We ended up with 67 stars, that are reported in Table \ref{Tabbasic}.

\subsection{Stellar parameters}


After radial velocity correction, the stellar spectrum is compared with the spectra of all stars in both libraries:\ Coelho05 and  MILES. 
The ETOILE code ranks all spectra from the library by similarity ($S$) to the target spectrum. 
$S$ is related to $\chi^{2}$, i.e., the most similar spectra have the smallest $S$ value
(for a definition of the similarity parameter, see Katz et al. 1998, and Dias et al. 2015).
A weighted mean of the stellar parameters T$_{\rm eff}$, log~$g$, [Fe/H], and [$\alpha$/Fe] of the most similar 
reference spectra is taken as the derived parameter of the target spectrum. The threshold to select the most
similar spectra is based on the normalized similarity, $S/S(1) \leq 1.1$ (Dias et al. 2015), applied
to results with both libraries.

The stellar parameters were first derived using the observed library MILES 
in the wavelength range of 4800-6000 {\rm \AA},
containing the MgI triplet lines, which is among the main
features commonly used in spectra of galaxies (Mg2, Mgb, Fe5270, Fe5335, Faber et al. 1985). 
From this procedure we obtained our first set of results.

Using the Coelho05 library, we carried out tests in different spectral regions, as well as with
the full spectral range of the MUSE spectra. 
As a check, we applied these calculations to spectra of the Sun,
Arcturus and the metal-rich red giant $\mu$ Leo (Lecureur et al. 2007). For the synthesis
of these spectra, the PFANT code (Barbuy et al. 2018b) was applied. The result indicated that the
most reliable region is 6000-6800 {\rm \AA}, which is, in fact, the region commonly used
to derive stellar parameters from high-resolution spectra (e.g. Barbuy et al. 2018c). This is explained by the following facts: it is widely known
that when bluer than 6000 {\rm \AA,} the continuum is progressively affected by molecular lines
as well as a large number of faint lines. 
When redder than 6800 {\rm \AA,} there are fewer lines, and, particularly fewer lines with well-defined oscillator strengths, 
and more numerous telluric lines.
The stellar parameters were then derived by running ETOILE with the library Coelho05
in the range 6000-6800 {\rm \AA}, obtaining a second set of results.

\begin{figure}[!htb]
\begin{center}
\includegraphics[width=3.7in]{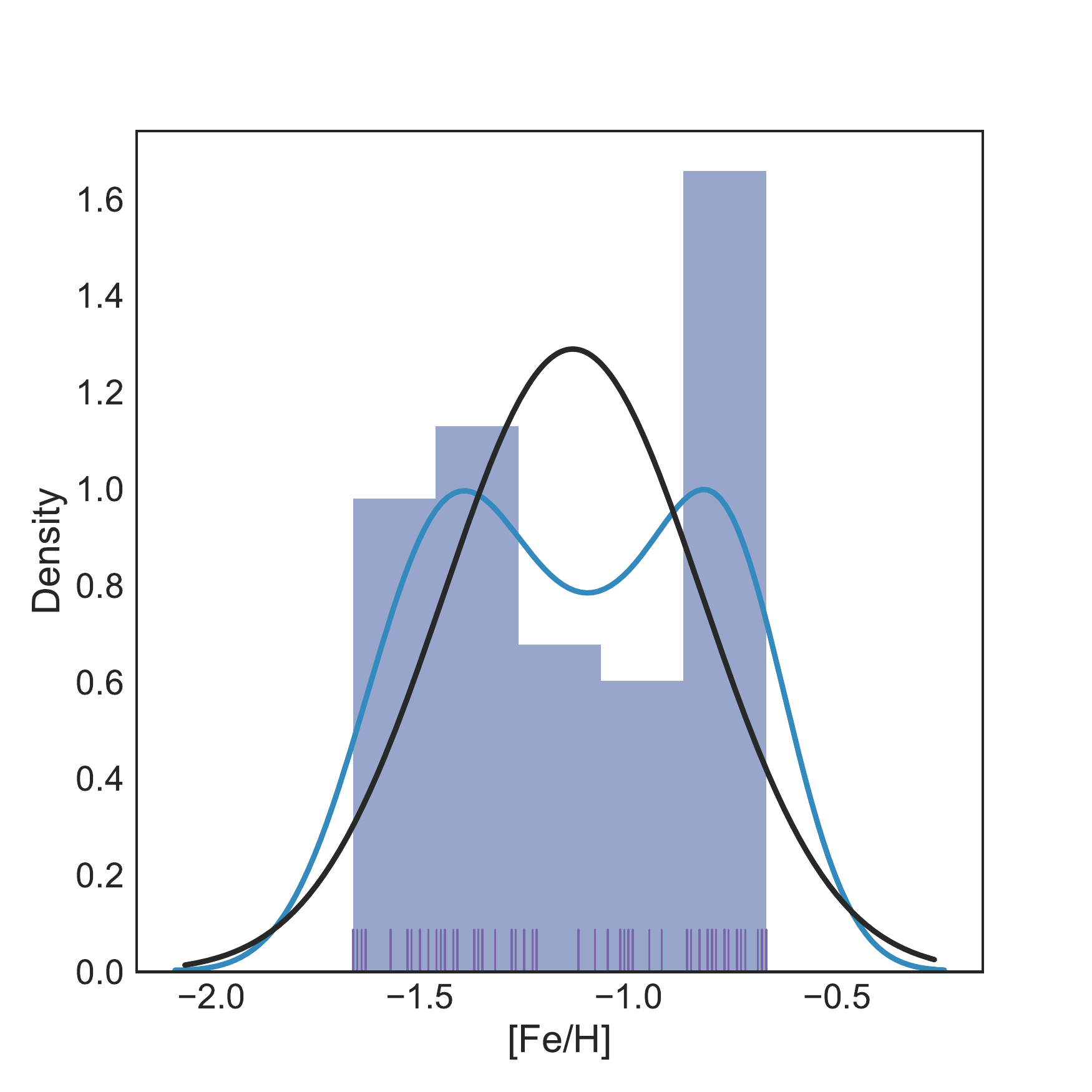}
\caption{ Metallicity distribution of sample stars  based on the optical analysis. The black curve represents a gaussian fit centered at a mean value of [Fe/H]=$-$1.12$\pm$0.12. The blue curve is a  KDE gaussian bandwidth estimated using Scott's rule}
\label{FeMean}
\end{center}
\end{figure}

From the final stellar parameters from the two applications (MILES and Coelho05), 
 a mean metallicity obtained from ETOILE along with the two libraries is [Fe/H]=-1.12$\pm$0.12, as shown in  Fig. \ref{FeMean}.
 It is important to note that there is a trend for lowering the metallicity
 as a function of lower S/N in this method. This is the reason
 for selecting only  high S/N$>$85 spectra; even so there is still a
 spread in metallicity values.

Finally, in Fig. \ref{fefinal} the metallicity distribution vs.
radial velocity distribution is shown, clearly indicating the locus of the cluster member stars.  There is no strong correlation between the possible
two peaks in metallicity hinted at in Fig. 11 and radial velocity,
 meaning that these are not two distinguished groups of  similar metallicity and radial
  velocity values.
In Fig. \ref{cornerplot}, the corner plot of different parameters of
the member stars is given.

\subsection{Uncertainties}
The uncertainties in this paper regarding the stellar parameters are the same as those that have already been described in section 3.2.2 in Dias et al. (2015). The uncertainties on the stellar parameters are computed using the average of squared residuals with the weighted $1/S^{2}$ as shown in the equation 

\begin{equation}
\sigma_{par(N)} = \sqrt[2]{\frac{1}{M}\frac{\sum_{m=1}^{M_{max}} (par_{m} - par)^{2} \times 1/S_{M}^{2} }{\sum_{m=1}^{M_{max}} 1/S_{n}^{2} }},
\end{equation}

\noindent where $par$ corresponds to the stellar parameters, T$_{\rm eff}$, log$g$, [Fe/H], and [$\alpha$/Fe], and N is the number of stars. The m, and M are counted as the number of the most similar stars in the library after the criteria of similarity S $\leq$ 1.1 is applied. 


\subsection{Metallicities from CaT}

We normalized the NIR portion of the spectra around the CaT lines in order to perform the techniques described in V\'asquez et al. (2015, 2018). The two stronger lines ($\lambda\lambda$ 8542, 8662 {\rm \AA}) were fitted using a combination of a Gaussian and a Lorentzian profile, and the equivalent widths were summed (W = W$_{8542} +$ W$_{8662}$). Since we used the same script as in V\'asquez et al. (2018), we were able to directly follow their calibrations, which we briefly describe here. The sum of the equivalent widths was first put into the same scale as Saviane et al. (2012) by applying the relation 

$$W_{S12} = 0.97\times W + 0.21.$$

The W$_{S12}$ was then corrected by gravity and temperature effects by applying the correction, resulting into the reduced equivalent width 

$$W' = W_{S12} + 0.55\times(V-V_{HB})$$

 where V$_{HB}$ = 20.35 mag (Ortolani et al. 1999). The W' was then converted into metallicity by applying the metallicity scale of Dias et al. (2016) represented by Eq. 5 of V\'asquez et al. (2018), that is, 

$$[Fe/H]_{\rm D16} = 0.055 \times W'^{2} + 0.13 \times W' - 2.68 $$

\begin{figure}[!ht]
\begin{center}
\includegraphics[width=3.3in]{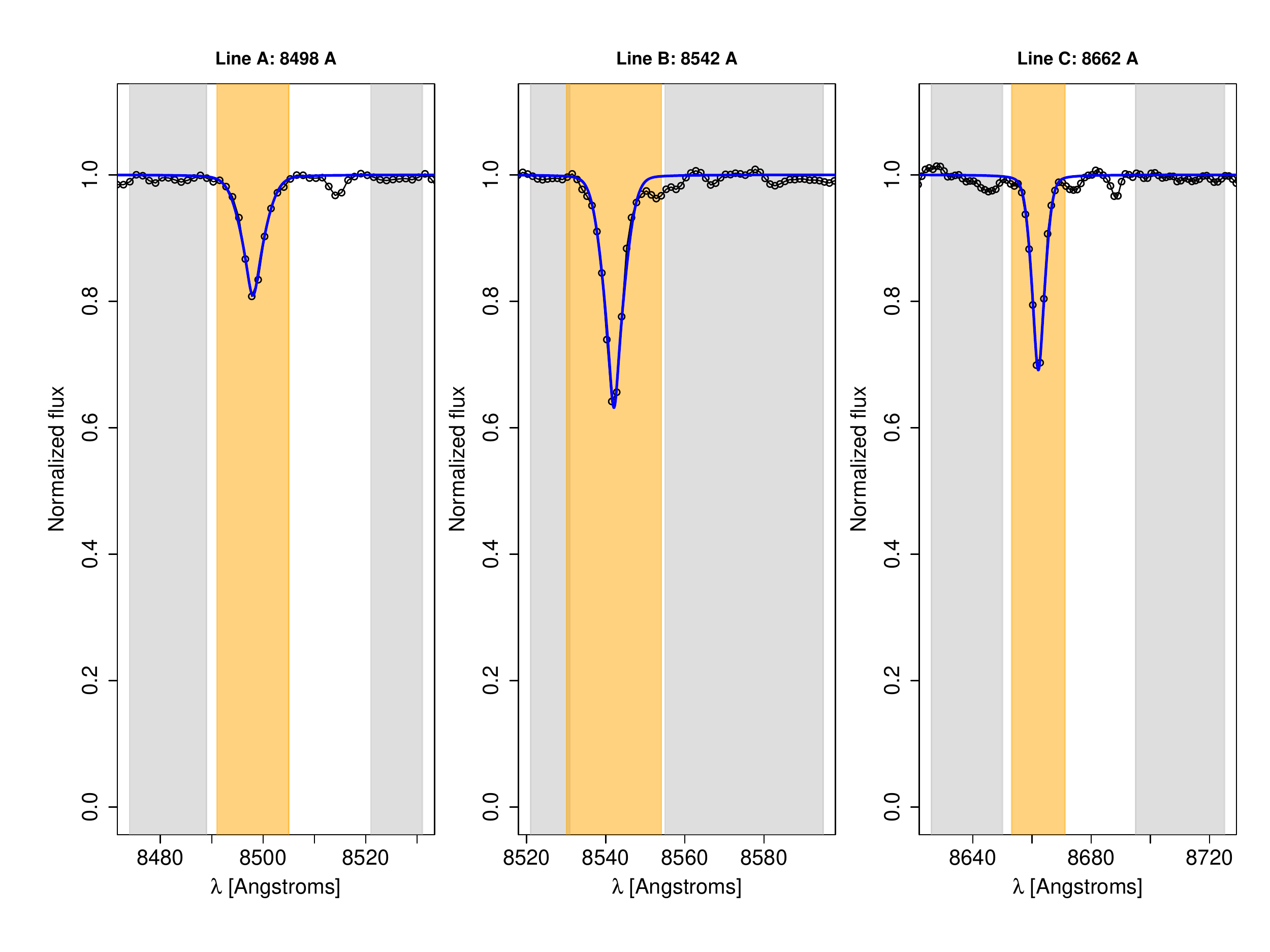}
\caption{Fit to CaT lines A: 8498 {\rm \AA}; B: 8542 {\rm \AA}, and C: 8662 {\rm \AA} for star 1378 as example.
  The shaded gray areas show the local continuum regions and the shaded orange areas show the line region defined
  by \cite{vasquez15}. The black lines and dots trace the observed spectrum in the rest frame and the blue lines
  are the  best model fit to the data, using a sum of Gaussian and Lorentzian functions.
  The spectrum has been locally normalized using the highlighted local continuum regions before
  the fitting. In this analysis we only use the sum of the equivalent widths of the two strongest
  lines (B+C) following the recipe of \cite{vasquez15,vasquez18}.}
\label{catspec}
\end{center}
\end{figure}

\begin{figure}[!ht]
\begin{center}
\includegraphics[width=3.7in]{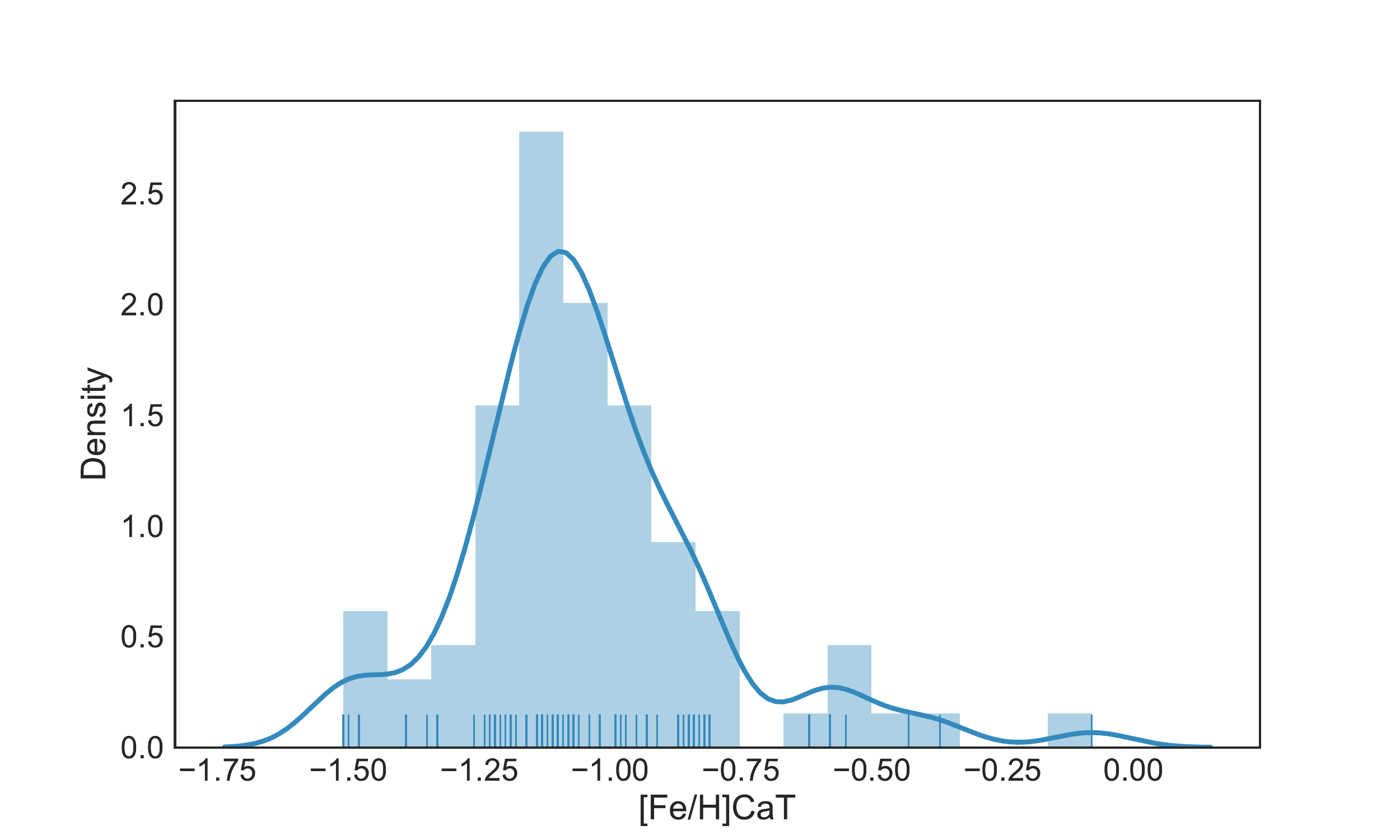}
\caption{ Metallicity distribution of sample stars  based on CaT analysis. }
\label{catdist}
\end{center}
\end{figure}

 Example of CaT lines are shown in Fig. \ref{catspec} for star 1378.
A typical error in  metallicity is of $\pm$ 0.1 dex.
The final list of cluster members where the metallicites derived from
procedures using the ETOILE code  and the CaT measurements are reported in Table \ref{Tabfinal}, which give a mean value of [Fe/H]=-1.09$\pm$0.15,  as shown in  Fig. \ref{catdist}.

\begin{figure}[!ht]
\begin{center}
\includegraphics[width=3.7in]{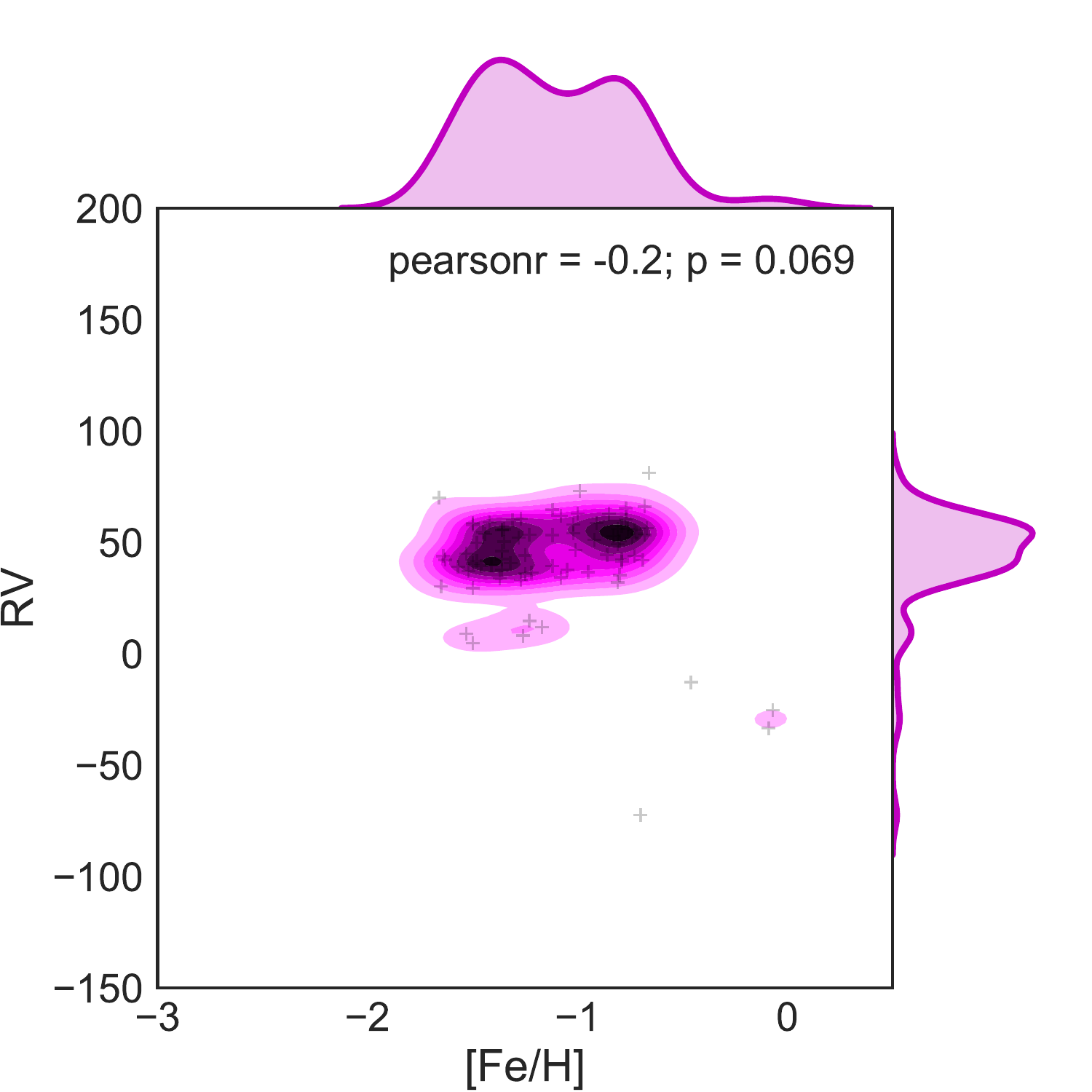}
\caption{ Metallicity distribution  based on optical analysis vs. radial velocity distribution, for identified member stars. }
\label{fefinal}
\end{center}
\end{figure}

\begin{figure*}[!htb]
\begin{center}
\includegraphics[width=6.1in]{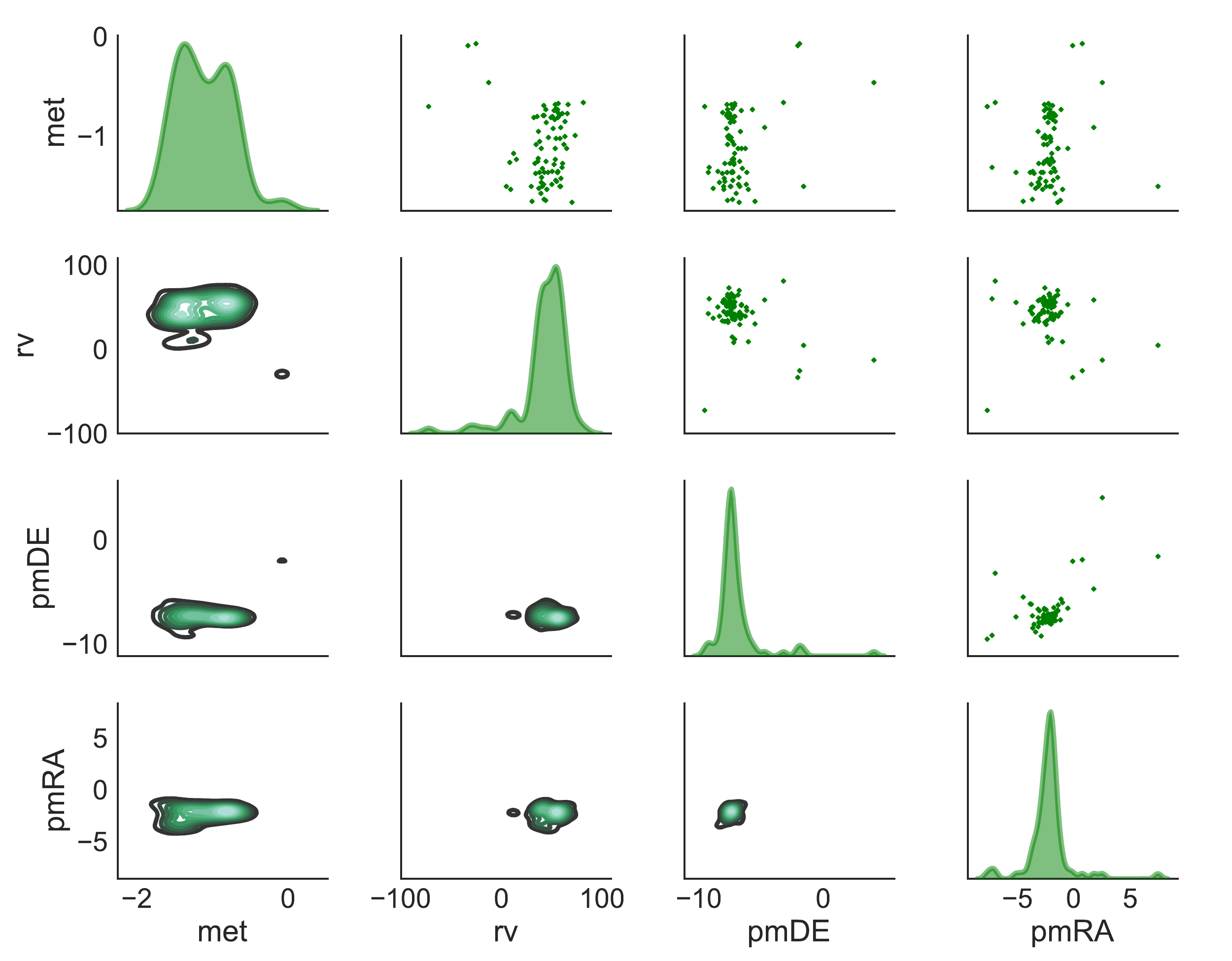}
\caption{Corner plot: metallicity, proper motion and radial velocities.}
\label{cornerplot}
\end{center}
\end{figure*}

 Finally, a comparison of metallicities for the same stars from the
ETOILE code and from CaT lines gives a mean difference of 
[Fe/H](ETOILE) - [Fe/H](CaT) $\approx$ -0.03 dex. In other words,
from ETOILE we get a mean of [Fe/H]=-1.12$\pm$0.12 and from CaT we get
[Fe/H]=-1.09$\pm$0.15, which are, therefore in excellent agreement.

Fig. \ref{fefinal} shows the metallicity distribution vs. the radial
velocity distribution for the identified 67 member stars.
Fig. \ref{cornerplot} shows a corner plot relating metallicities,
proper motions, and radial velocities.

\subsection{Color-magnitude diagrams of member stars}

\begin{figure}[!htb]
\begin{center}
\includegraphics[width=3.7in]{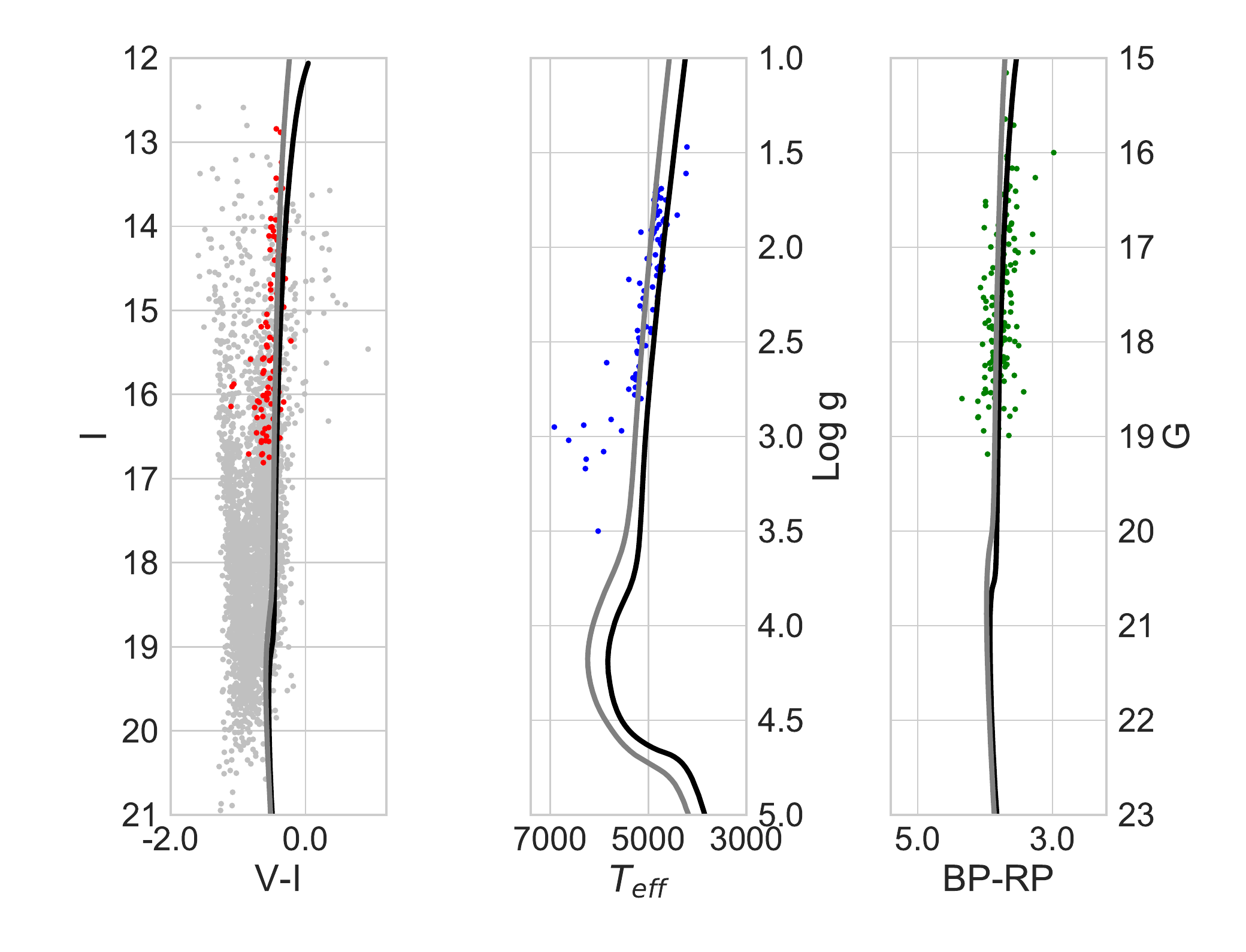}
\caption{I vs. V-I color-magnitude diagram showing all stars (gray) and member stars (red)
(left panel), compared with the log g vs. T$_{\rm eff}$ diagram (middle panel), and 
CMD in Gaia magnitudes and colors G vs. BP-RP for the stars in common
(right panel). Dartmouth isochrones of 13 Gyr, and [Fe/H]=-1.0 are overplotted in black and isochrones of 13 Gyr, and [Fe/H]=-2.0 are overplotted in gray.}
\label{HRD}
\end{center}
\end{figure}

In Fig. \ref{HRD} we compare the I vs. V-I color-magnitude diagram showing all stars
where the member stars are highlighted, and the resulting
 log g vs. T$_{\rm eff}$ diagram.  At the RGB base, a small
 trend towards high temperatures might be present.
  The brighter the  RGB stars, the closer the isochrones
   get to the more metal-rich ones, again indicating that the metallicity
 is not bimodal and that the spread is due to S/N effects.
 On the right panel, member stars identified in the Gaia survey, 
 are plotted with Gaia colors G vs. BR-RP.
 Dartmouth isochrones of 13 Gyr,  [Fe/H]=-2.0 and [Fe/H]=-1.0 are overplotted.
 The I values were corrected by A$_{\rm I}$ cf. 
 Schlafly \& Finkbeiner (2011)\footnote{\\https://irsa.ipac.caltech.edu/applications/DUST};
 for the Gaia magnitudes no corrections were applied.
 In this Fig. we clearly see the RGB stars.  A BHB appears more clearly
 present in Fig. \ref{HRD}b, confirming ealier evidence by Ortolani et al.
 (1999).

\section{Discussion}

 Photometric data indicate a broad range  of metallicities: 
from V vs. V-I CMDs, Ortolani et al. (1999) deduced a metallicity
of [Fe/H]$\sim$-2.0, Valenti et al. (2007)
instead derived [Fe/H]$\sim$-1.2 from K$_{s}$ vs. J-K$_{s}$ CMDs; Bica et al. (1998) derived [Z/Z$_{\odot}$]=-1.01 using integrated spectra
of CaT lines. 
 High-resolution spectroscopic analyses based on CaT lines from
the literature are available: Armandroff \& Zinn (1998) obtained
[Fe/H]=-0.99, and from 6 stars,
 V\'asquez et al. (2018) report Fe/H]=-1.08$\pm$0.14, -1.21$\pm$0.15, -1.16$\pm$0.21,
on the scales of Dias et al. (2016), Saviane et al. (2012), and their own.
 The compilation by
Harris (1996, 2010 edition) reports [Fe/H]= -1.05, 
whereas average metallicity compiled by Carretta et al. (2009),
adopting a value from Harris (1996) from before 2010, was given as
[Fe/H]=-2.07$\pm$0.09.
In the present work, the metallicity derived from the 67 selected member stars turned out to be of
 [Fe/H]=-1.12$\pm$0.12 from the optical and 
 [Fe/H]=-1.09$\pm$0.15 from CaT lines, therefore, a final metallicity of
 [Fe/H]=-1.10$\pm$0.15 was adopted.

 The radial velocity of our sample stars was
double-checked with synthetic spectra exhaustively, therefore, we suggest
that our value of v$_{r}^{h}$ = 58.1$\pm$1.1 km s$^{-1}$ is more robust 
 than the higher value of v$_{\rm r}$ = 71.4 km s$^{-1}$, given in V\'asquez et al. (2018), due to the higher numbers of stars.  

Terzan 9 is now included in  the list of moderately metal-poor globular clusters with a BHB 
similar to HP~1 (Barbuy et al. 2016), NGC 6522 (Barbuy et al. 2014),
and NGC 6558 (Barbuy et al. 2018c).
Therefore, Terzan 4 continues,  
so far, to be the most metal-poor cluster in the Galactic bulge, with [Fe/H]=-1.6 (Origlia \& Rich 2004).
 Other potential bulge clusters with metallicities 
below that of Terzan~4, and within 3.5 kpc from the
Galactic center, specifically NGC 6144, NGC 6273, NGC 6287, NGC 6293,
NGC 6293, NGC 6333, NGC 6541, are classified as halo intruders in Bica et al. (2016). The orbital classification by P\'erez-Villegas et al. (2019) determines these clusters as inner/outer halo, thick disk or disk, and none of them are classified as a bulge member. As for NGC 6681,
it has a radial velocity of 216.62 km/s, and apogalactic distance
of 4.97 kpc (Baumgardt et al. 2019), which might indicate that it is a
 halo intruder. 

 Terzan 9 has a blue HB, but not an extended one (see Ortolani et al. 1999). The moderately metal-poor
 metallicity found for Terzan 9 
correspond essentially to the lower end of the metallicity
distribution of the bulk bulge stellar population. As a matter of fact,
due to a fast chemical enrichment in the Galactic bulge, such as
the one modeled by e.g. Cescutti et al. (2008), the iron abundance
of [Fe/H]$\sim$-1.3 is reached very fast, and stellar populations start
to form in more significant numbers from there on, as confirmed
by metallicity distribution functions (MDF) given in 
Zoccali et al. (2008, 2017),
Hill et al. (2011), Ness et al. (2013),
Rojas-Arriagada et al. (2014, 2017) - see also Barbuy et al. (2018a).

\begin{figure}[!htb]
\begin{center}
\includegraphics[width=3.7in]{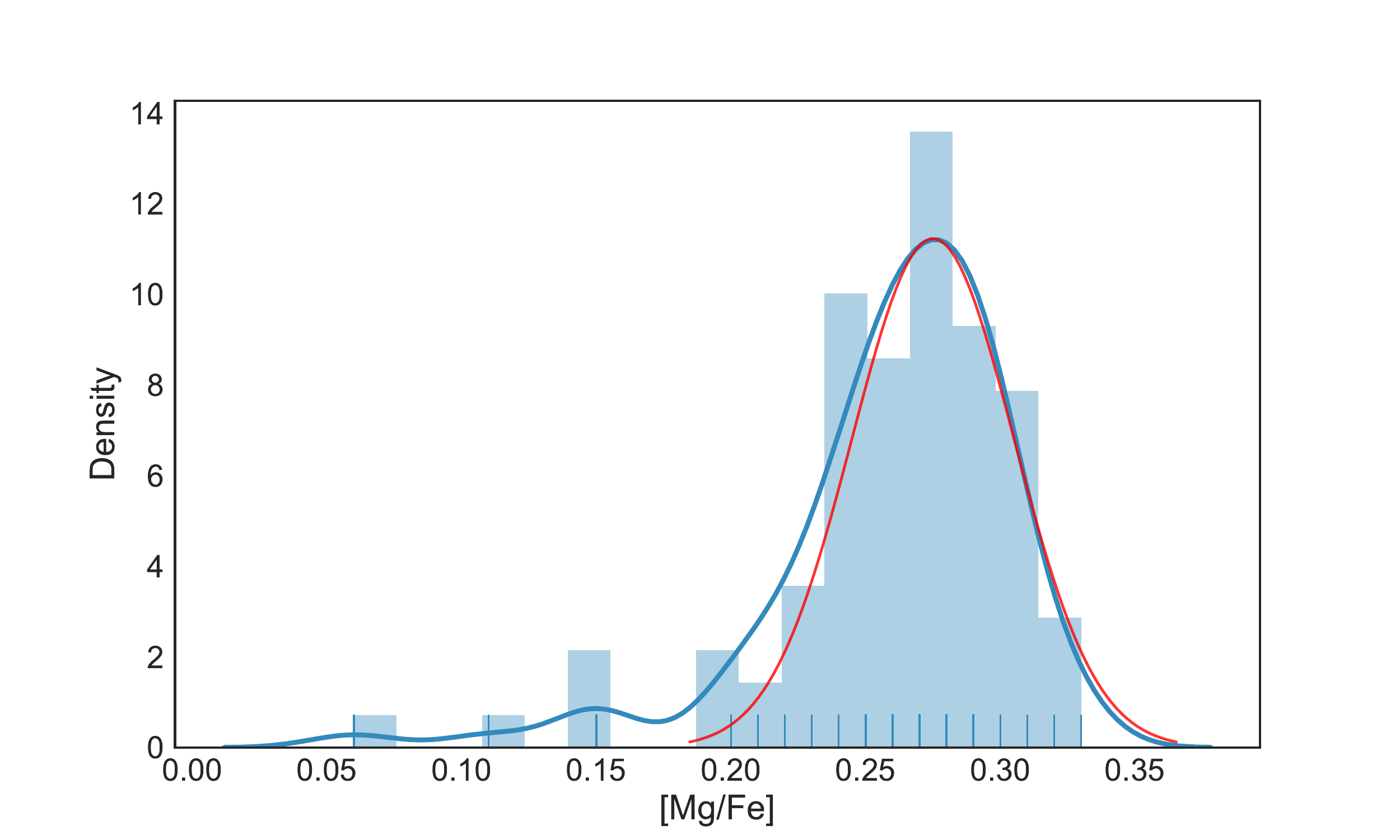}
\caption{Distribution in [Mg/Fe]. A KDE plot indicates a mean value of [Mg/Fe]=+0.27. }
\label{distalpha}
\end{center}
\end{figure}

The derivation of
 Mg-to-iron is based on the fitting of the MgI triplet lines
(see Dias et al. 2015, 2016).
In Fig. \ref{distalpha}, the distribution of enhancement in the $\alpha$-element Mg is shown with a mean value of  [Mg/Fe]=+0.27$\pm$0.03.
The sigma of the distribution results is also $\pm$0.03.
This enhancement is similar to those reported
in the Galactic bulge by Barbuy et al. (2018a) and Schultheis et al. (2017).
This indicates that the stars in Terzan 9 were formed from gas
resulting from an early fast chemical enrichment by core-collapse supernovae.

\section{Conclusions}

We obtained MUSE datacubes for the bulge compact globular cluster Terzan~9.
Using the software pampelMUSE by Kamann et al. (2013, 2018), we were able to extract the
spectra of over 600 stars. The sample was reduced to 67 member stars by selecting
spectra with S/N$>$85 and with compatible radial velocities and proper motions.
These spectra were analysed based on a full spectrum fitting with the ETOILE code in the area of
4600-5600 {\rm \AA}, compared with 
 a grid of observed spectra (MILES, S\'anchez-Bl\'asquez et al. 2006). In the area of 6000-6800 {\rm \AA}, they were compared 
 with a grid of synthetic spectra by Coelho et al. (2005, Coelh05).
 The CaT lines were also measured in order to obtain an independent
 derivation of metallicity. Both methods give very close mean results, with an adopted mean of [Fe/H]=-1.10$\pm$0.15. This mean value is the outcome of the
 combination of a range of values where, in particular with regard to the optical region,
 two metallicity peaks are seen. In order
 to confirm metallicities, further observations with high resolution
 spectroscopy are of great interest. The present paper allows for a
 reliable target selection for such studies.
 
We were able to derive a mean heliocentric 
radial velocity of v$^{\rm h}_{\rm r}$ = 58.1$\pm$1.1 {\rm km s}$^{-1}$, which
is somewhat lower than the value from V\'asquez et al. (2018) based on 6 stars,
but the values are in agreement within uncertainties.
These metallicities place Terzan 9 as a new member of the
moderately metal-poor clusters with a blue horizontal branch that are found in the
Galactic bulge.

\begin{acknowledgements}
We are grateful to David Katz for helpful comments on the Etoile code, and to
Angeles P\'erez-Villegas for help with the proper motion and orbital information.
HE, BB, EB and EC acknowledge partial financial support, 
grants and fellowships from FAPESP, CNPq and CAPES - Finance code 001.
HE and EC are grateful for their visit to ESO 
in Santiago, within ESO's office for science programme,
under supervision of B. Dias.
\end{acknowledgements}



\begin{appendix}

\section{Extracted stars from the MUSE datacubes}

\scalefont{0.7}
\begin{longtable}{c|c|c|c|c|c|c|c|c|c}
\caption{\label{Tabbasic}Identified member stars from MUSE datacubes selected with S/N$>$85. 
Columns correspond to: ID from NTT 2012 data, coordinates (RA,DEC-J2000), proper motions from Gaia, NTT pixels x, y, NTT V, NTT V-I, and S/N.
}\\  \\

\hline
\hline
ID & RA (J2000)  & DEC (J2000) & pmRA & pmDEC & X & Y & V & V-I & S/N \\
\hline
0072     & 270.41522393678054 & -26.83960326833801 & -1.6470 & -7.1760 & 1114.09 &  991.63 &   17.87 &   3.812 & 129.80 \\ 
0081     & 270.40828978697527 & -26.83592435071913 & -1.8160 & -7.1100 &  929.55 & 1101.25 &   18.42 &   3.841 & 125.00 \\ 
0084     & 270.41578613193025 & -26.83405140131479 & -2.5750 & -7.6500 & 1129.81 & 1157.05 &   18.74 &   3.937 & 114.13 \\ 
0089     & 270.39921802548110 & -26.83204463546964 & -2.1770 & -7.3860 &  687.86 & 1217.57 &   18.58 &   3.717 & 114.03 \\ 
0092     & 270.40840223869230 & -26.83177706400694 & -2.0010 & -7.4560 &  932.48 & 1225.10 &   18.88 &   4.252 & 121.02 \\ 
0437     & 270.40307930558885 & -26.84742893173680 & -1.8330 & -7.4950 &  790.31 &  757.66 &   17.94 &   3.820 & 106.96 \\ 
0473     & 270.41432441417822 & -26.84421846865514 & -3.0010 & -7.9650 & 1090.16 &  853.74 &   19.68 &   3.417 &  99.41 \\ 
0513     & 270.40034269441361 & -26.84074036414967 & -2.2920 & -7.4720 &  717.94 &  957.50 &   18.12 &   3.950 & 117.32 \\ 
0520     & 270.41170073374064 & -26.84023870564067 & -2.4930 & -7.5020 & 1020.38 &  972.22 &   17.97 &   3.688 & 129.40 \\ 
0549     & 270.41867198860143 & -26.83655980866259 & -2.0500 & -7.5950 & 1206.10 & 1082.43 &   18.34 &   3.984 & 122.03 \\ 
0554     & 270.39678117444549 & -26.83615846721820 & -2.1590 & -7.5650 &  622.57 & 1094.94 &   18.25 &   3.849 & 113.81 \\ 
0571     & 270.41484913724622 & -26.83505477091640 & -1.7740 & -7.5350 & 1104.23 & 1127.62 &   17.69 &   4.144 & 116.83 \\ 
0578     & 270.41814729715378 & -26.83448619589934 & -2.6740 & -7.1830 & 1192.40 & 1144.92 &   18.47 &   4.122 & 124.17 \\ 
0582     & 270.40105497443244 & -26.83408484711128 & -2.0170 & -8.0870 &  736.12 & 1156.14 &   17.82 &   3.900 & 104.30 \\ 
0584     & 270.41260029354157 & -26.83395106386613 & -1.4360 & -6.8510 & 1044.23 & 1160.69 &   19.74 &   4.040 & 122.81 \\ 
0595     & 270.41147584179771 & -26.83294768449528 & -3.1640 & -8.3520 & 1014.50 & 1190.66 &   20.21 &   3.861 & 101.16 \\ 
0610     & 270.40585328417666 & -26.83164327803886 & -2.6140 & -7.7440 &  864.91 & 1229.44 &   18.27 &   3.985 & 116.47 \\ 
0611     & 270.40315427957711 & -26.83147604535684 & -2.6990 & -7.5050 &  792.25 & 1234.08 &   19.50 &   3.688 & 108.72 \\ 
0615     & 270.41642328041041 & -26.83107468591398 & -2.2590 & -7.1160 & 1146.62 & 1246.95 &   18.63 &   4.089 & 120.39 \\ 
0631     & 270.42020856001682 & -26.82940233961694 & -1.4360 & -6.7630 & 1247.46 & 1296.56 &   18.68 &   3.530 &  88.71 \\ 
0645     & 270.41293762517887 & -26.82746238703481 & -2.4270 & -7.2970 & 1053.09 & 1354.89 &   19.53 &   3.807 &  85.38 \\ 
0936     & 270.40296684444041 & -26.83475376096855 & -2.1910 & -7.6940 &  787.64 & 1136.18 &   17.35 &   4.109 & 108.45 \\ 
1342     & 270.41282518149904 & -26.84361649668046 & -3.1550 & -7.4820 & 1050.16 &  871.39 &   19.54 &   3.358 & 112.85 \\ 
1353     & 270.40570334248764 & -26.84298107834307 & -2.0320 & -7.2650 &  860.67 &  890.05 &   19.27 &   3.669 & 125.42 \\ 
1368     & 270.40915191171359 & -26.84181056153792 & -2.0550 & -7.3130 &  952.73 &  925.10 &   17.73 &   3.714 & 140.45 \\ 
1380     & 270.40821481905323 & -26.84090758315861 & -2.3690 & -7.6940 &  927.83 &  952.88 &   18.55 &   3.358 & 122.89 \\ 
1399     & 270.40349166142846 & -26.83963671249548 & -2.4830 & -7.9150 &  801.46 &  990.74 &   19.76 &   3.821 & 109.47 \\ 
1406     & 270.41121346685679 & -26.83936915895905 & -2.6190 & -6.5930 & 1007.24 &  998.86 &   17.34 &   3.912 & 100.73 \\ 
1413     & 270.40401647407731 & -26.83886749437426 & -1.6550 & -7.3100 &  815.57 & 1013.86 &   20.19 &   3.636 &  97.36 \\ 
1414     & 270.41739773041394 & -26.83880060559504 & -1.8270 & -7.2750 & 1172.89 & 1015.09 &   19.36 &   3.793 & 127.92 \\ 
1426     & 270.40802739886061 & -26.83793104786989 & -1.6750 & -7.0070 &  922.82 & 1041.32 &   19.62 &   3.630 & 110.36 \\ 
1433     & 270.41754764447046 & -26.83752971128347 & -3.0530 & -6.8120 & 1176.21 & 1053.57 &   19.48 &   3.204 & 112.48 \\ 
1441     & 270.41147584179771 & -26.83692770373768 & -1.8900 & -7.1330 & 1014.86 & 1071.40 &   17.49 &   3.923 & 117.52 \\ 
1447     & 270.42043342100408 & -26.83645947343477 & -2.1240 & -7.1800 & 1253.94 & 1085.56 &   19.00 &   3.681 &  95.24 \\ 
1448     & 270.41151332384356 & -26.83572367904923 & -4.4920 & -5.4910 & 1015.41 & 1107.86 &   20.09 &   3.801 & 101.22 \\ 
1514     & 270.40776500966109 & -26.83184395693179 & -1.6220 & -7.7590 &  915.49 & 1223.40 &   19.15 &   4.193 & 120.53 \\ 
1545     & 270.40735268158420 & -26.83003783411370 & -2.7280 & -7.4070 &  904.67 & 1277.56 &   19.82 &   4.459 &  97.37 \\ 
2009     & 270.40877707631000 & -26.84067347647692 & -3.6420 & -8.4780 &  942.11 &  959.88 &   20.09 &   3.377 &  85.56 \\ 
2011     & 270.41027640464006 & -26.84037248146045 & -2.5990 & -6.7240 &  982.66 &  968.88 &   18.39 &   3.704 & 134.41 \\ 
2013     & 270.40559088598837 & -26.84033903752032 & -0.5410 & -6.5740 &  857.80 &  969.23 &   18.46 &   3.706 & 131.43 \\ 
2032     & 270.41080116118627 & -26.83816516022189 & -2.3340 & -7.1630 &  996.01 & 1034.08 &   17.77 &   3.760 & 129.99 \\ 
2034     & 270.41559873410057 & -26.83803138179427 & -1.8210 & -7.2890 & 1124.13 & 1038.38 &   18.78 &   3.591 & 130.93 \\ 
2038     & 270.40971416066765 & -26.83689425878022 & -2.7190 & -7.7350 &  967.22 & 1072.12 &   19.62 &   3.568 & 110.74 \\ 
2055     & 270.40086753309708 & -26.83562334308101 & -1.8830 & -7.3890 &  731.65 & 1110.99 &   18.01 &   3.885 & 117.00 \\ 
2062     & 270.40836475480876 & -26.83448619589934 & -3.1760 & -6.6610 &  931.93 & 1144.49 &   20.25 &   4.076 &  97.77 \\ 
2352     & 270.41053878345559 & -26.84568994221631 & -3.3940 & -8.8530 &  989.78 &  809.38 &   19.38 &   3.284 & 109.77 \\ 
2355     & 270.41379968677040 & -26.84502109297108 & -2.1480 & -8.0110 & 1076.29 &  829.32 &   19.03 &   3.459 & 136.97 \\ 
2518     & 270.41589857036234 & -26.84053970101285 & -1.4380 & -6.7830 & 1132.82 &  963.10 &   20.05 &   3.554 &  89.15 \\ 
2523     & 270.40405396052904 & -26.84010492966281 & -2.7670 & -7.9990 &  816.13 &  976.28 &   19.84 &   3.724 & 113.03 \\ 
2525     & 270.40802739886061 & -26.83993770946819 & -3.5260 & -8.1130 &  922.68 &  981.08 &   18.97 &   3.557 & 130.86 \\ 
2561     & 270.41196310666669 & -26.83468686976047 & -3.7640 & -6.1900 & 1027.36 & 1138.66 &   20.40 &   3.656 &  90.63 \\ 
2562     & 270.40157980722722 & -26.83461997851290 & -1.9920 & -7.2260 &  750.53 & 1140.22 &   19.18 &   3.456 & 106.00 \\ 
2573     & 270.41233792325039 & -26.83318180714274 & -1.7830 & -6.5100 & 1037.44 & 1183.96 &   20.57 &   4.048 &  86.81 \\ 
2575     & 270.40971416066765 & -26.83234565261057 & -3.1110 & -7.9170 &  967.65 & 1208.54 &   19.98 &   4.012 & 109.40 \\ 
2748     & 270.41245036779372 & -26.83956982417065 & -3.8680 & -6.1480 & 1040.01 &  992.71 &   18.99 &   3.412 & 121.33 \\ 
2848     & 270.40881455994997 & -26.83679392384860 & -1.3960 & -6.2670 &  943.93 & 1075.06 &   20.03 &   3.633 &  92.35 \\ 
2851     & 270.40990157587862 & -26.83318180714274 & -2.1280 & -7.6950 &  972.15 & 1183.44 &   18.23 &   4.281 & 112.67 \\ 
2903     & 270.41121346685679 & -26.84221188294944 & -3.6720 & -7.2970 & 1007.51 &  913.48 &   19.93 &   3.361 &  92.13 \\ 
2963     & 270.41147584179771 & -26.83498787988597 & -2.0390 & -7.2200 & 1014.40 & 1129.59 &   19.16 &   3.956 & 133.76 \\ 
2968     & 270.40202966045496 & -26.83181051047429 & -2.1270 & -7.6880 &  762.65 & 1224.70 &   19.53 &   3.612 & 108.05 \\ 
3042     & 270.40862714152860 & -26.84211155272999 & -2.4570 & -6.7110 &  938.05 &  916.28 &   20.07 &   3.516 &  94.47 \\ 
3068     & 270.40937681189303 & -26.83361660506254 & -1.1350 & -5.7000 &  958.32 & 1170.29 &   18.39 &   4.245 & 129.01 \\ 
3088     & 270.41462425646279 & -26.83752971128347 & -1.9910 & -7.4270 & 1098.60 & 1053.20 &   19.63 &   3.565 & 115.52 \\ 
3104     & 270.41346235971645 & -26.83773037975444 & -1.1950 & -7.6990 & 1067.02 & 1047.56 &   19.44 &   3.429 & 104.55 \\ 
3162     & 270.40746513496180 & -26.84254631637180 & -2.5720 & -7.5880 &  907.93 &  903.03 &   16.76 &   3.917 &  99.54 \\ 
3187     & 270.41020143906485 & -26.83769693503397 & -1.9390 & -7.7700 &  980.88 & 1048.00 &   17.63 &   3.717 & 110.81 \\ 
3194     & 270.41275021893512 & -26.84047281322153 & -1.7490 & -7.7720 & 1048.33 &  965.88 &   16.95 &   4.072 & 102.53 \\ 
\hline
\hline

\end{longtable}

\scalefont{0.5}
\begin{longtable}{c|c|c|c|c|c|c|c|c|c}
\caption{\label{Tabfinal}Identified members stars from MUSE datacubes selected with S/N$>$85. 
Columns correspond to: ID and results from the present work: v$_{r}^{h}$ (km s$^{-1}$), T$_{\rm eff}$ (K), log g, [Fe/H], [Mg/Fe], and [Fe/H]$_{\rm CaT}$.} \\

\hline
ID & v$_r$ (km s$^{-1}$)  & T$_{\rm eff}$ (K)  & log g & [Fe/H] & [Mg/Fe] & EWa & EWb & EW' & [Fe/H]$_{CaT}$ \\
\hline
0072     &   63.06 & 4660 $\pm$ 42   &  1.99 $\pm$ 0.15 &  -0.85 $\pm$ 0.07 &   0.27 $\pm$ 0.03 &  3.095  $\pm$   0.114  &  2.365 $\pm$  0.102 &  5.460 $\pm$  0.153 &  -1.20 $\pm$   0.09 \\ 
0081     &   35.38 & 4893 $\pm$ 49   &  2.21 $\pm$ 0.13 &  -0.80 $\pm$ 0.06 &   0.24 $\pm$ 0.02 &  2.994  $\pm$   0.161  &  2.383 $\pm$  0.141 &  5.377 $\pm$  0.214 &  -1.07 $\pm$   0.12 \\ 
0084     &   32.17 & 4885 $\pm$ 52   &  2.33 $\pm$ 0.13 &  -0.81 $\pm$ 0.07 &   0.24 $\pm$ 0.02 &  2.963  $\pm$   0.138  &  2.240 $\pm$  0.121 &  5.203 $\pm$  0.183 &  -1.06 $\pm$   0.11 \\ 
0089     &   66.10 & 4988 $\pm$ 28   &  2.72 $\pm$ 0.10 &  -0.68 $\pm$ 0.05 &   0.21 $\pm$ 0.02 &  3.626  $\pm$   0.138  &  3.135 $\pm$  0.161 &  6.761 $\pm$  0.212 &  -0.08 $\pm$   0.16 \\ 
0092     &   49.20 & 4788 $\pm$ 46   &  2.15 $\pm$ 0.12 &  -0.81 $\pm$ 0.07 &   0.26 $\pm$ 0.02 &  3.128  $\pm$   0.143  &  2.410 $\pm$  0.126 &  5.539 $\pm$  0.190 &  -0.81 $\pm$   0.12 \\ 
0437     &   57.47 & 4628 $\pm$ 36   &  2.12 $\pm$ 0.12 &  -0.79 $\pm$ 0.06 &   0.27 $\pm$ 0.03 &  3.198  $\pm$   0.112  &  2.411 $\pm$  0.096 &  5.610 $\pm$  0.148 &  -1.09 $\pm$   0.09 \\ 
0473     &   59.01 & 5220 $\pm$ 90   &  2.19 $\pm$ 0.17 &  -1.42 $\pm$ 0.10 &   0.30 $\pm$ 0.03 &  2.489  $\pm$   0.143  &  2.040 $\pm$  0.118 &  4.529 $\pm$  0.185 &  -1.14 $\pm$   0.10 \\ 
0513     &   44.50 & 4630 $\pm$ 37   &  2.06 $\pm$ 0.12 &  -0.86 $\pm$ 0.06 &   0.29 $\pm$ 0.03 &  3.144  $\pm$   0.139  &  2.293 $\pm$  0.117 &  5.437 $\pm$  0.182 &  -1.13 $\pm$   0.10 \\ 
0520     &   58.13 & 4670 $\pm$ 66   &  1.69 $\pm$ 0.20 &  -1.02 $\pm$ 0.10 &   0.29 $\pm$ 0.03 &  3.121  $\pm$   0.140  &  2.389 $\pm$  0.125 &  5.510 $\pm$  0.188 &  -1.14 $\pm$   0.11 \\ 
0549     &   54.18 & 4728 $\pm$ 49   &  1.88 $\pm$ 0.18 &  -1.02 $\pm$ 0.08 &   0.30 $\pm$ 0.03 &  3.081  $\pm$   0.118  &  2.390 $\pm$  0.106 &  5.471 $\pm$  0.159 &  -1.04 $\pm$   0.10 \\ 
0554     &   56.09 & 4934 $\pm$ 39   &  2.45 $\pm$ 0.12 &  -0.81 $\pm$ 0.06 &   0.24 $\pm$ 0.02 &  3.154  $\pm$   0.132  &  2.647 $\pm$  0.119 &  5.801 $\pm$  0.178 &  -0.87 $\pm$   0.11 \\ 
0571     &   60.37 & 4524 $\pm$ 49   &  1.88 $\pm$ 0.17 &  -0.77 $\pm$ 0.08 &   0.29 $\pm$ 0.03 &  3.213  $\pm$   0.132  &  2.499 $\pm$  0.121 &  5.712 $\pm$  0.179 &  -1.11 $\pm$   0.10 \\ 
0578     &   61.99 & 4554 $\pm$ 62   &  1.75 $\pm$ 0.21 &  -1.08 $\pm$ 0.09 &   0.30 $\pm$ 0.03 &  3.071  $\pm$   0.150  &  2.361 $\pm$  0.143 &  5.432 $\pm$  0.207 &  -1.02 $\pm$   0.12 \\ 
0582     &   56.05 & 4632 $\pm$ 44   &  1.94 $\pm$ 0.14 &  -0.76 $\pm$ 0.07 &   0.30 $\pm$ 0.03 &  3.252  $\pm$   0.147  &  2.509 $\pm$  0.138 &  5.761 $\pm$  0.201 &  -1.04 $\pm$   0.12 \\ 
0584     &   64.72 & 4972 $\pm$ 87   &  2.09 $\pm$ 0.22 &  -1.12 $\pm$ 0.10 &   0.28 $\pm$ 0.03 &  2.797  $\pm$   0.155  &  2.199 $\pm$  0.136 &  4.996 $\pm$  0.206 &  -0.84 $\pm$   0.13 \\ 
0595     &   39.55 & 5286 $\pm$ 39   &  2.74 $\pm$ 0.12 &  -1.46 $\pm$ 0.06 &   0.28 $\pm$ 0.01 & ---  $\pm$  ---  & --- $\pm$ --- & --- $\pm$ --- &  --- $\pm$  --- \\ 
0610     &   53.41 & 4767 $\pm$ 36   &  2.26 $\pm$ 0.12 &  -0.68 $\pm$ 0.06 &   0.24 $\pm$ 0.02 &  3.185  $\pm$   0.144  &  2.404 $\pm$  0.127 &  5.589 $\pm$  0.192 &  -0.99 $\pm$   0.11 \\ 
0611     &   60.67 & 5134 $\pm$ 78   &  2.27 $\pm$ 0.16 &  -1.27 $\pm$ 0.08 &   0.27 $\pm$ 0.03 &  2.698  $\pm$   0.124  &  2.021 $\pm$  0.097 &  4.719 $\pm$  0.158 &  -1.09 $\pm$   0.10 \\ 
0615     &   56.39 & 5056 $\pm$ 46   &  2.42 $\pm$ 0.12 &  -0.67 $\pm$ 0.06 &   0.24 $\pm$ 0.02 &  3.080  $\pm$   0.106  &  2.377 $\pm$  0.094 &  5.457 $\pm$  0.141 &  -0.95 $\pm$   0.09 \\ 
0631     &   70.03 & 5348 $\pm$ 43   &  2.78 $\pm$ 0.13 &  -1.66 $\pm$ 0.06 &   0.29 $\pm$ 0.01 &  2.523  $\pm$   0.098  &  2.209 $\pm$  0.230 &  4.732 $\pm$  0.250 &  -1.35 $\pm$   0.12 \\ 
0645     &   54.05 & 5684 $\pm$ 42   &  2.97 $\pm$ 0.09 &  -0.72 $\pm$ 0.05 &   0.15 $\pm$ 0.02 &  2.890  $\pm$   0.127  &  2.281 $\pm$  0.125 &  5.171 $\pm$  0.178 &  -0.81 $\pm$   0.11 \\ 
0936     &   65.37 & 4264 $\pm$ 33   &  1.83 $\pm$ 0.14 &  -0.77 $\pm$ 0.07 &   0.24 $\pm$ 0.03 &  3.325  $\pm$   0.103  &  2.633 $\pm$  0.087 &  5.958 $\pm$  0.135 &  -1.08 $\pm$   0.09 \\ 
1342     &   57.98 & 4821 $\pm$ 64   &  1.71 $\pm$ 0.19 &  -1.50 $\pm$ 0.10 &   0.30 $\pm$ 0.04 &  2.483  $\pm$   0.117  &  1.919 $\pm$  0.099 &  4.403 $\pm$  0.153 &  -1.26 $\pm$   0.09 \\ 
1353     &   37.79 & 4714 $\pm$ 134  &  1.81 $\pm$ 0.34 &  -1.36 $\pm$ 0.15 &   0.26 $\pm$ 0.05 &  2.740  $\pm$   0.160  &  2.098 $\pm$  0.161 &  4.837 $\pm$  0.227 &  -1.10 $\pm$   0.12 \\ 
1368     &   37.91 & 4589 $\pm$ 70   &  1.85 $\pm$ 0.21 &  -1.05 $\pm$ 0.10 &   0.28 $\pm$ 0.04 &  2.859  $\pm$   0.146  &  2.154 $\pm$  0.117 &  5.013 $\pm$  0.188 &  -1.48 $\pm$   0.09 \\ 
1380     &   34.38 & 5498 $\pm$ 410  &  2.17 $\pm$ 0.38 &  -1.08 $\pm$ 0.25 &   0.20 $\pm$ 0.07 &  2.628  $\pm$   0.152  &  2.368 $\pm$  0.161 &  4.996 $\pm$  0.222 &  -1.24 $\pm$   0.12 \\ 
1399     &   54.69 & 4796 $\pm$ 104  &  1.90 $\pm$ 0.27 &  -1.45 $\pm$ 0.12 &   0.29 $\pm$ 0.03 &  2.688  $\pm$   0.134  &  2.092 $\pm$  0.119 &  4.781 $\pm$  0.179 &  -0.97 $\pm$   0.11 \\ 
1406     &   52.26 & 4582 $\pm$ 62   &  1.87 $\pm$ 0.17 &  -0.74 $\pm$ 0.07 &   0.32 $\pm$ 0.04 &  3.298  $\pm$   0.123  &  2.660 $\pm$  0.127 &  5.958 $\pm$  0.176 &  -1.08 $\pm$   0.10 \\ 
1413     &   56.02 & 5110 $\pm$ 78   &  2.33 $\pm$ 0.18 &  -1.35 $\pm$ 0.09 &   0.27 $\pm$ 0.03 &  2.430  $\pm$   0.165  &  1.879 $\pm$  0.161 &  4.309 $\pm$  0.231 &  -1.10 $\pm$   0.13 \\ 
1414     &   53.88 & 4788 $\pm$ 96   &  1.72 $\pm$ 0.27 &  -1.44 $\pm$ 0.14 &   0.27 $\pm$ 0.04 &  2.774  $\pm$   0.129  &  2.200 $\pm$  0.116 &  4.974 $\pm$  0.173 &  -0.99 $\pm$   0.11 \\ 
1426     &   39.18 & 4753 $\pm$ 76   &  1.70 $\pm$ 0.22 &  -1.57 $\pm$ 0.12 &   0.29 $\pm$ 0.04 &  2.650  $\pm$   0.193  &  2.095 $\pm$  0.213 &  4.745 $\pm$  0.287 &  -1.04 $\pm$   0.16 \\ 
1433     &   41.82 & 6586 $\pm$ 96   &  3.12 $\pm$ 0.11 &  -1.48 $\pm$ 0.08 &   0.24 $\pm$ 0.03 & ---  $\pm$  ---  & --- $\pm$ --- & --- $\pm$ --- &  --- $\pm$  --- \\ 
1441     &   42.03 & 4622 $\pm$ 44   &  1.86 $\pm$ 0.12 &  -0.69 $\pm$ 0.07 &   0.33 $\pm$ 0.03 &  3.430  $\pm$   0.162  &  2.703 $\pm$  0.136 &  6.134 $\pm$  0.212 &  -0.93 $\pm$   0.13 \\ 
1447     &   52.27 & 5217 $\pm$ 49   &  2.50 $\pm$ 0.12 &  -1.28 $\pm$ 0.06 &   0.27 $\pm$ 0.02 &  2.828  $\pm$   0.129  &  2.288 $\pm$  0.162 &  5.116 $\pm$  0.207 &  -1.02 $\pm$   0.12 \\ 
1448     &   30.36 & 4865 $\pm$ 76   &  1.85 $\pm$ 0.24 &  -1.65 $\pm$ 0.15 &   0.29 $\pm$ 0.05 & ---  $\pm$  ---  & --- $\pm$ --- & --- $\pm$ --- &  --- $\pm$  --- \\ 
1514     &   41.44 & 5244 $\pm$ 106  &  2.48 $\pm$ 0.18 &  -0.79 $\pm$ 0.10 &   0.20 $\pm$ 0.04 &  3.005  $\pm$   0.183  &  2.317 $\pm$  0.180 &  5.322 $\pm$  0.256 &  -0.85 $\pm$   0.15 \\ 
1545     &   46.67 & 5072 $\pm$ 42   &  2.52 $\pm$ 0.12 &  -1.01 $\pm$ 0.07 &   0.25 $\pm$ 0.02 &  3.139  $\pm$   0.115  &  2.511 $\pm$  0.111 &  5.650 $\pm$  0.160 &  -0.37 $\pm$   0.12 \\ 
2009     &   50.38 & 5277 $\pm$ 54   &  2.44 $\pm$ 0.14 &  -1.35 $\pm$ 0.07 &   0.28 $\pm$ 0.02 &  2.466  $\pm$   0.142  &  1.906 $\pm$  0.149 &  4.371 $\pm$  0.206 &  -1.10 $\pm$   0.12 \\ 
2011     &   44.21 & 4776 $\pm$ 78   &  1.73 $\pm$ 0.22 &  -1.25 $\pm$ 0.15 &   0.29 $\pm$ 0.04 &  2.972  $\pm$   0.135  &  2.457 $\pm$  0.122 &  5.429 $\pm$  0.182 &  -1.04 $\pm$   0.11 \\ 
2013     &   53.35 & 4859 $\pm$ 96   &  1.75 $\pm$ 0.25 &  -1.12 $\pm$ 0.15 &   0.28 $\pm$ 0.04 &  2.861  $\pm$   0.147  &  2.269 $\pm$  0.148 &  5.130 $\pm$  0.209 &  -1.20 $\pm$   0.11 \\ 
2032     &   50.53 & 4774 $\pm$ 45   &  2.11 $\pm$ 0.12 &  -0.80 $\pm$ 0.07 &   0.26 $\pm$ 0.02 &  3.166  $\pm$   0.147  &  2.525 $\pm$  0.135 &  5.692 $\pm$  0.199 &  -1.10 $\pm$   0.11 \\ 
2034     &   35.99 & 4848 $\pm$ 96   &  1.79 $\pm$ 0.25 &  -1.25 $\pm$ 0.15 &   0.29 $\pm$ 0.04 &  2.781  $\pm$   0.183  &  2.256 $\pm$  0.167 &  5.037 $\pm$  0.248 &  -1.14 $\pm$   0.13 \\ 
2038     &   55.53 & 5277 $\pm$ 57   &  2.56 $\pm$ 0.14 &  -1.36 $\pm$ 0.07 &   0.27 $\pm$ 0.02 &  2.278  $\pm$   0.152  &  1.607 $\pm$  0.138 &  3.885 $\pm$  0.205 &  -1.51 $\pm$   0.10 \\ 
2055     &   52.87 & 4690 $\pm$ 40   &  1.98 $\pm$ 0.14 &  -0.83 $\pm$ 0.07 &   0.28 $\pm$ 0.03 &  3.165  $\pm$   0.132  &  2.521 $\pm$  0.121 &  5.687 $\pm$  0.179 &  -1.02 $\pm$   0.11 \\ 
2062     &   36.72 & 5252 $\pm$ 42   &  2.69 $\pm$ 0.12 &  -0.95 $\pm$ 0.06 &   0.22 $\pm$ 0.02 &  2.671  $\pm$   0.156  &  2.060 $\pm$  0.139 &  4.731 $\pm$  0.209 &  -0.83 $\pm$   0.13 \\ 
2352     &   37.09 & 5190 $\pm$ 136  &  1.92 $\pm$ 0.17 &  -1.52 $\pm$ 0.10 &   0.32 $\pm$ 0.03 &  2.542  $\pm$   0.145  &  2.512 $\pm$  0.165 &  5.054 $\pm$  0.220 &  -0.93 $\pm$   0.13 \\ 
2355     &   58.71 & 4854 $\pm$ 86   &  1.92 $\pm$ 0.28 &  -1.50 $\pm$ 0.18 &   0.24 $\pm$ 0.04 &  2.329  $\pm$   0.145  &  1.886 $\pm$  0.130 &  4.215 $\pm$  0.195 &  -1.51 $\pm$   0.09 \\ 
2518     &   39.85 & 5290 $\pm$ 56   &  2.55 $\pm$ 0.14 &  -1.41 $\pm$ 0.07 &   0.27 $\pm$ 0.02 & ---  $\pm$  ---  & --- $\pm$ --- & --- $\pm$ --- &  --- $\pm$  --- \\ 
2523     &   45.03 & 4810 $\pm$ 56   &  1.75 $\pm$ 0.17 &  -1.53 $\pm$ 0.09 &   0.30 $\pm$ 0.03 &  2.585  $\pm$   0.172  &  2.110 $\pm$  0.172 &  4.695 $\pm$  0.244 &  -0.99 $\pm$   0.14 \\ 
2525     &   33.89 & 4820 $\pm$ 134  &  1.79 $\pm$ 0.35 &  -1.37 $\pm$ 0.14 &   0.26 $\pm$ 0.05 &  2.644  $\pm$   0.143  &  1.959 $\pm$  0.120 &  4.603 $\pm$  0.186 &  -1.33 $\pm$   0.10 \\ 
2561     &   50.00 & 5950 $\pm$ 56   &  2.91 $\pm$ 0.11 &  -1.48 $\pm$ 0.07 &   0.27 $\pm$ 0.02 &  2.508  $\pm$   0.149  &  2.516 $\pm$  0.179 &  5.024 $\pm$  0.233 &  -0.58 $\pm$   0.15 \\ 
2562     &   36.49 & 4922 $\pm$ 108  &  1.95 $\pm$ 0.29 &  -1.22 $\pm$ 0.15 &   0.25 $\pm$ 0.04 &  2.827  $\pm$   0.137  &  2.258 $\pm$  0.127 &  5.085 $\pm$  0.186 &  -0.98 $\pm$   0.11 \\ 
2573     &   40.69 & 5317 $\pm$ 38   &  2.67 $\pm$ 0.11 &  -1.32 $\pm$ 0.06 &   0.26 $\pm$ 0.02 & ---  $\pm$  ---  & --- $\pm$ --- & --- $\pm$ --- &  --- $\pm$  --- \\ 
2575     &   33.24 & 5331 $\pm$ 42   &  2.74 $\pm$ 0.12 &  -1.27 $\pm$ 0.06 &   0.26 $\pm$ 0.02 &  2.451  $\pm$   0.214  &  1.962 $\pm$  0.261 &  4.413 $\pm$  0.337 &  -1.11 $\pm$   0.17 \\ 
2748     &   46.19 & 4775 $\pm$ 134  &  1.83 $\pm$ 0.34 &  -1.36 $\pm$ 0.15 &   0.26 $\pm$ 0.05 &  2.800  $\pm$   0.170  &  2.240 $\pm$  0.201 &  5.040 $\pm$  0.264 &  -1.07 $\pm$   0.14 \\ 
2848     &   39.52 & 5335 $\pm$ 40   &  2.78 $\pm$ 0.12 &  -1.12 $\pm$ 0.06 &   0.22 $\pm$ 0.02 & ---  $\pm$  ---  & --- $\pm$ --- & --- $\pm$ --- &  --- $\pm$  --- \\ 
2851     &   55.62 & 4632 $\pm$ 38   &  2.10 $\pm$ 0.12 &  -0.73 $\pm$ 0.07 &   0.25 $\pm$ 0.02 &  3.299  $\pm$   0.129  &  2.523 $\pm$  0.107 &  5.821 $\pm$  0.168 &  -0.86 $\pm$   0.11 \\ 
2903     &   42.28 & 4760 $\pm$ 55   &  1.73 $\pm$ 0.18 &  -1.63 $\pm$ 0.10 &   0.30 $\pm$ 0.03 &  2.335  $\pm$   0.132  &  1.933 $\pm$  0.123 &  4.267 $\pm$  0.180 &  -1.21 $\pm$   0.10 \\ 
2963     &   53.44 & 4802 $\pm$ 135  &  1.78 $\pm$ 0.35 &  -1.23 $\pm$ 0.15 &   0.26 $\pm$ 0.05 &  2.850  $\pm$   0.113  &  2.142 $\pm$  0.102 &  4.992 $\pm$  0.152 &  -1.04 $\pm$   0.09 \\ 
2968     &   62.98 & 5318 $\pm$ 40   &  2.70 $\pm$ 0.12 &  -1.00 $\pm$ 0.05 &   0.24 $\pm$ 0.02 &  2.539  $\pm$   0.114  &  2.052 $\pm$  0.103 &  4.590 $\pm$  0.154 &  -1.16 $\pm$   0.09 \\ 
3042     &   29.55 & 4926 $\pm$ 86   &  1.96 $\pm$ 0.27 &  -1.50 $\pm$ 0.19 &   0.25 $\pm$ 0.04 &  2.584  $\pm$   0.174  &  2.113 $\pm$  0.177 &  4.697 $\pm$  0.248 &  -0.91 $\pm$   0.14 \\ 
3068     &   44.04 & 4716 $\pm$ 38   &  2.10 $\pm$ 0.12 &  -0.73 $\pm$ 0.06 &   0.27 $\pm$ 0.03 &  3.216  $\pm$   0.143  &  2.584 $\pm$  0.132 &  5.800 $\pm$  0.194 &  -0.82 $\pm$   0.12 \\ 
3088     &   40.64 & 4887 $\pm$ 86   &  1.93 $\pm$ 0.28 &  -1.50 $\pm$ 0.19 &   0.25 $\pm$ 0.04 &  2.437  $\pm$   0.172  &  1.977 $\pm$  0.161 &  4.414 $\pm$  0.235 &  -1.23 $\pm$   0.12 \\ 
3104     &   43.91 & 4927 $\pm$ 75   &  1.91 $\pm$ 0.24 &  -1.64 $\pm$ 0.15 &   0.29 $\pm$ 0.05 &  2.212  $\pm$   0.150  &  1.802 $\pm$  0.148 &  4.014 $\pm$  0.211 &  -1.50 $\pm$   0.10 \\ 
3162     &   73.13 & 4016 $\pm$ 34   &  1.47 $\pm$ 0.17 &  -0.99 $\pm$ 0.12 &   0.28 $\pm$ 0.04 &  3.385  $\pm$   0.099  &  2.663 $\pm$  0.085 &  6.048 $\pm$  0.131 &  -1.22 $\pm$   0.08 \\ 
3187     &   42.36 & 4646 $\pm$ 44   &  1.95 $\pm$ 0.14 &  -0.79 $\pm$ 0.07 &   0.29 $\pm$ 0.03 &  3.203  $\pm$   0.132  &  2.506 $\pm$  0.109 &  5.709 $\pm$  0.172 &  -1.13 $\pm$   0.10 \\ 
3194     &   51.23 & 4040 $\pm$ 34   &  1.61 $\pm$ 0.18 &  -0.92 $\pm$ 0.12 &   0.29 $\pm$ 0.04 &  3.356  $\pm$   0.101  &  2.582 $\pm$  0.088 &  5.938 $\pm$  0.134 &  -1.22 $\pm$   0.08 \\ 
\hline
\hline

\end{longtable}

\end{appendix}

\end{document}